\documentclass[acmsmall,screen]{acmart}
\usepackage{graphicx}
\usepackage[most]{tcolorbox}
\usepackage{textcomp}
\usepackage{xspace}
\usepackage{multirow}
\usepackage{rotating}
\usepackage{tikz}
\usepackage{soul}
\usepackage[normalem]{ulem}
\usepackage{tabularx}
\usepackage{makecell}
\usepackage{url}
\usepackage{enumitem}
\usepackage{colortbl}
\usepackage{booktabs}
\usepackage{pifont}
\usepackage{ifthen}
\usepackage[caption=false]{subfig}
\usepackage[table]{xcolor}
\usepackage[export]{adjustbox}
\usepackage{amsmath,amsfonts}
\usepackage{listings}
\usepackage{algorithmic}
\usepackage{comment}
\usepackage{ifxetex}

\usepackage{tikz,xcolor}
\definecolor{iconPurple}{HTML}{6A2FE8}
\definecolor{iconGreen}{HTML}{2EA043}

\newcommand{\reductionicon}[1]{%
\begin{tikzpicture}[baseline={(current bounding box.south)}]
  \begin{scope}[x=1ex,y=1ex,line join=round,line cap=round]
    \pgfmathsetmacro{\H}{2.0}
    \pgfmathsetmacro{\W}{0.45}
    \pgfmathsetmacro{\S}{0.28}
    \pgfmathsetmacro{\p}{#1/100*\H}

    \fill[iconPurple] (0,0) rectangle (\W,\H);

    \fill[iconGreen] ({\W+\S},0) rectangle ({2*\W+\S},\p);

    \draw[black,opacity=.25,line width=.08ex] (0,0) rectangle (\W,\H);
    \draw[black,opacity=.25,line width=.08ex] ({\W+\S},0) rectangle ({2*\W+\S},\H);
  \end{scope}
\end{tikzpicture}%
}

\ifxetex
  \usepackage{fontspec}
  \usepackage{xeCJK}
\fi

\def\BibTeX{{\rm B\kern-.05em{\sc i\kern-.025em b}\kern-.08em
    T\kern-.1667em\lower.7ex\hbox{E}\kern-.125emX}}

\definecolor{darkgreen}{rgb}{0.0, 0.5, 0.0}

\newcommand{\ie}{\emph{i.e.,}\xspace}
\newcommand{\eg}{\emph{e.g.,}\xspace}
\newcommand{\etc}{etc.\xspace}
\newcommand{\etal}{\emph{et~al.}\xspace}
\newcommand{\secref}[1]{Section~\ref{#1}\xspace}

\newcommand{\figref}[1]{Fig.~\ref{#1}\xspace}

\newcommand{\tabref}[1]{Table~\ref{#1}\xspace}

\tcbset{
  on line,
  boxsep=1pt,
  left=2pt,
  right=2pt,
  top=1pt,
  bottom=1pt,
  colframe=blue!70!black, 
  colback=blue!70!black,  
  coltext=white,          
  rounded corners,
  boxrule=0pt             
}
\newcommand{\royalval}[1]{\tcbox{#1}}

\newcommand{\sidep}{\textit{$SIDE_{py}$}\xspace}
\newcommand{\side}{\textit{SIDE}\xspace}
\newcommand{\ccpair}{$\langle \mbox{\it code}, \mbox{\it comment} \rangle$ }
\renewcommand{\rq}[1]{RQ\textsubscript{#1}}
\newcommand{\cbleuApproach}{\textit{CrystalBLEU-guided optimization}\xspace}
\newcommand{\astApproach}{\textit{AST-based optimization}\xspace}
\newcommand{\functionApproach}{\textit{FS-based optimization}\xspace}

\newboolean{showcomments}
\setboolean{showcomments}{true}

\ifthenelse{\boolean{showcomments}}{
  \newcommand{\nb}[2]{\fbox{\bfseries\sffamily\scriptsize#1}
    {\sf\small$\blacktriangleright$\textit{#2}$\blacktriangleleft$}
  }
}{
  \newcommand{\nb}[2]{}
}

\newtcolorbox{promptbox}{colback=white, arc=0.5mm, top=1mm, bottom=1mm, left=1mm, right=1mm, title=System prompt used for generation}
\newtcolorbox{resultbox}{colback=white, arc=0.5mm, top=1mm, bottom=1mm, left=1mm, right=1mm}

\AtBeginDocument{\providecommand\BibTeX{{Bib\TeX}}}



\begin{document}
\title{Not All Tokens Matter: Data-Centric Optimization for Efficient Code Summarization}


\author{Saima Afrin}
\email{safrin@wm.edu}
\affiliation{%
	\institution{AURA @ Dept. of Computer Science, William \& Mary}
	\country{USA}
}

\author{Zaiyu Cheng}
\email{zcheng06@wm.edu}
\affiliation{%
	\institution{AURA @ Dept. of Computer Science, William \& Mary}
	\country{USA}
}

\author{Tushar Sharma}
\email{tushar@dal.ca}
\affiliation{%
	\institution{Dalhousie University}
	\country{CA}
}

\author{Alexander Serebrenik}
\email{a.serebrenik@tue.nl}
\affiliation{%
	\institution{Eindhoven University of Technology}
	\country{NL}
}

\author{Massimiliano Di Penta}
\email{dipenta@unisannio.it}
\affiliation{%
	\institution{Dept. of Engineering, University of Sannio}
	\country{IT}
}

\author{Antonio Mastropaolo}
\email{amastropaolo@wm.edu}
\affiliation{%
	\institution{AURA @ Dept. of Computer Science, William \& Mary}
	\country{USA}
}

\begin{abstract}
The rapid advancement of Large Language Models (LLMs) has revolutionized software engineering automation, particularly in automated code summarization, which enhances program comprehension and supports development activities. However, training LLMs for code summarization remains computationally expensive, with performance deteriorating on longer inputs--challenges that intensify when handling millions of code–comment pairs. We investigate strategic data optimization through targeted token reduction to minimize computational overhead while maintaining summary quality. We compare three token-level reduction techniques--(i) Abstract Syntax Tree (AST) representations, (ii) Function Signatures, and (iii) CrystalBLEU-guided pruning--combined with semantic filtering, evaluating them on Java and Python in standalone and cascaded reduction settings. Our findings reveal highly language-dependent optimal strategies: AST-based optimization achieves 37\% performance improvements in Java with 56-73\% token reduction but shows up to 49\% degradation in Python. Conversely, Function Signatures perform poorly in Java but optimally in Python, achieving 83\% token reduction while maintaining quality. CrystalBLEU demonstrates cross-language robustness with 60-72\% reduction. These results challenge assumptions about cross-language transferability, demonstrating that which tokens are kept matters more than how many are removed, making language-aware token curation essential for efficient code summarization.

\end{abstract}
\maketitle

\section{Introduction} \label{sec:introduction}

Automated code summarization seeks to generate a Natural Language (NL) description of software artifacts, for purposes such as (re)documentation or supporting code comprehension \cite{zhang2024review, arthur2020automatic, wang2023codet5p}.
In this context, Large Language Models (LLMs) have shown remarkable capability to generate natural-language summaries from source code (code-to-NL) by processing code as sequences of tokens. However, this success hinges on the availability of high-performance hardware that can efficiently handle the vast number of tokens and parameters required for model training and inference. 

Hardware accelerators---such as Graphics Processing Units (GPUs), Tensor Processing Units (TPUs), or custom AI chips---play a critical role in meeting the computational demands of modern LLMs. Nevertheless, relying solely on hardware advancements is neither scalable nor sustainable. Recent studies warn that the marginal gains from brute-force hardware scaling are diminishing, and the environmental cost of ever-larger models continues to rise~\cite{velasco2025toward, mastropaolo2025path}. 
In addition to computational concerns, 
training datasets must be vast, diverse, and richly annotated--especially in domains like code intelligence \cite{puri2021codenet}.
Scaling dataset size typically takes two forms: increasing the number of training examples or expanding the token footprint of individual inputs \cite{gu2025effectiveness,tang2023biocoder}. Both approaches result in a large volume of tokens being processed during training, which amplifies memory requirements, energy consumption, and training time \cite{lozhkov2024starcoder}.

Code summarization datasets constructed from mined software repositories---even when subjected to basic quality control---inevitably remain noisy, often containing numerous low-quality $\langle code, comment \rangle$ pairs, as noted in prior work \cite{BirdBADBFD09,HerzigJZ13,BachmannBRDB10,mastropaolo2024evaluating}.
Such noise poses a serious challenge: \textit{low-quality instances can cap the model's ability to generalize, introduce spurious patterns during training, and inflate computational cost without corresponding gains in performance \cite{wang2024empirical}.}

To improve code summarization,
Shi \etal~\cite{shi2022we} introduced CAT, a heuristic-driven method that cleans datasets by removing structurally defective examples such as trivial methods or summaries that simply mirror code. While 
filtering surface-level flaws, CAT overlooks deeper semantic inconsistencies between code and comments. Vitale \etal~\cite{vitale2025optimizing} observed that many pairs passing CAT remain misaligned, often because developers neglect to update documentation when code changes (cf. Steidl  \etal~\cite{Steidl:icpc2013}). These misalignments weaken supervision, confuse models during training, and inflate computational costs. To address this, Vitale \etal employed the SIDE metric \cite{mastropaolo2024evaluating}, which evaluates semantic alignment and prunes noisy examples. Their results revealed that removing up to half of the data preserved performance, underscoring that dataset quality outweighs sheer size.

This shift from quantity to quality motivates our own hypothesis: if dataset-level pruning already improves learning, then finer-grained control at the token level may yield even greater benefits. In particular, \ul{\textit{we argue that not all tokens contribute equally to the learning process\textemdash many are syntactically necessary but semantically uninformative for summarization. 
These low-utility tokens dilute meaningful training signals, inflate computational costs, and hinder the model's ability to generate concise and insightful summaries.}}

To address the challenge of low-utility tokens, we propose token-level data optimization methods that identify and retain only the most informative content while discarding redundant or semantically weak tokens. These techniques leverage syntactical source code manipulation and equivalent code representations to transform the original input before feeding it to the model. Our goal is to strategically reduce training token volume, enhancing model efficiency while potentially improving summary quality. We present three distinct strategies: \textbf{(i)} \cbleuApproach, \textbf{(ii)} \functionApproach, and \textbf{(iii)} \astApproach.

Our results reveal that integrating token-level reduction techniques with existing instance-level filtering methods into a unified framework yields significant efficiency improvements, achieving 55-83\% token reduction and 13-33\% entropy reduction across different optimization strategies. However, optimal strategies are highly language-dependent: \astApproach achieves strong performance improvements in Java but shows substantial degradation in Python, while \functionApproach exhibits the opposite pattern, emerging as optimal for Python despite its poor performance in Java. \cbleuApproach demonstrates cross-language robustness, consistently delivering 60-72\% token reduction with 22-33\% entropy reduction across both programming languages. These substantial compression gains come at varying performance trade-offs depending on language and dataset characteristics, with some configurations maintaining quality and others showing degradation. This indicates that token-level optimization strategies must be calibrated to language-specific characteristics. This finding highlights the potential of token-level reduction as a crucial yet underexplored dimension of data-centric optimization for code summarization.

In summary, we make the following contributions:

\begin{itemize}
    \item A comprehensive empirical evaluation of three token-level optimization strategies (AST, Function Signatures, CrystalBLEU) across Java and Python, demonstrating language-specific effectiveness patterns.
    
    \item Development and validation of \sidep, a Python-specific semantic alignment metric showing substantial improvement over language-agnostic baselines in correlation with human adequacy judgments.

    \item A new high-quality Python evaluation benchmark--\textit{PyBench}, consisting of 500 manually curated code-summary pairs from recent GitHub repositories, with human annotations across three quality dimensions to enable reliable evaluation of code summarization models.
    
    
    
    \item A complete replication package with datasets, models, benchmarks, and analysis scripts to support reproducible research in data-centric code summarization \cite{replication}.
\end{itemize}
\section{Background and Related Work}
\label{sec:related}

This section discusses advances related to evaluation metrics for code summarization, as well as their application in data optimization. We do not discuss further work on code summarization in general, which is extensively analyzed in a systematic literature review by Zhu \etal~\cite{zhang2024review}.




\setlength{\textfloatsep}{10pt}

\subsection{Lexical-based Evaluation Metrics}
\label{sub:Lexical-based}


Evaluating code summaries is non-trivial. Metrics such as BLEU~\cite{papineni:acl2002}, ROUGE~\cite{lin:tsbo2004}, and METEOR~\cite{banerjee:acl2005} are popular for their simplicity and reproducibility, offering surface-level similarity assessments commonly used in natural language tasks.
Below we list the metrics used in our study.

\textbf{BLEU (Bilingual Evaluation Understudy)}~\cite{papineni:acl2002} quantifies the n-gram overlap between a generated and a reference summary, assigning a similarity score between 0 (no overlap) and 1 (perfect match). BLEU can be computed at the sentence level for various values of \emph{n} (specifically, \emph{n} = 1, 2, 3, and 4), capturing different granularities of matching sequences. 
\textbf{METEOR (Metric for Evaluation of Translation with Explicit ORdering)}~\cite{banerjee:acl2005} improves upon BLEU by incorporating both stemming and synonym matching, allowing for greater flexibility in capturing semantically similar but lexically distinct summaries. It computes the harmonic mean of unigram precision and recall, with a higher weight placed on recall. Like BLEU, METEOR produces a score between 0 and 1, with higher values indicating closer alignment to the reference summary.
%
\textbf{ROUGE (Recall-Oriented Understudy for Gisting Evaluation)}~\cite{lin:tsbo2004} evaluates summaries through n-gram overlap with references. Variants include ROUGE-N (n-gram matching), ROUGE-L (longest common subsequence), and ROUGE-W (weighted ROUGE-L), reporting precision, recall, and F1-scores.
%
%
\textbf{chrF}~\cite{popovic:wmt2015} operates at the character level, unlike word-based metrics above.

Finally, \textbf{c\_coeff}~\cite{Steidl:icpc2013} measures summary-code similarity by calculating the percentage of summary words matching code words (using Levenshtein distance~\cite{levenshtein:dp1966} < 2). 

These metrics are popular, but fail with semantically equivalent, lexically different descriptions.




\subsection{Semantic-based Evaluation Metrics}
\label{sub:Semantic-based}
To better capture semantic meaning, embedding-based metrics have emerged. These metrics utilize pre-trained language models to produce contextualized vector representations of text, allowing for more nuanced comparisons between generated and reference summaries. For instance,

\textbf{BERTScore}~\cite{zhang2019bertscore} computes the similarity between generated and reference summaries using embeddings derived from the BERT model~\cite{devlin2018bert}, which is pre-trained on large-scale English text corpora. It reports three components---precision (BERTScore-P), recall (BERTScore-R), and F1-score (BERTScore-F1)---that collectively reflect the degree of semantic alignment between the summaries.

\textbf{Sentence-BERT}~\cite{reimers2019sentence} enhances the original BERT architecture by introducing a Siamese network structure to produce fixed-size sentence embeddings optimized for semantic similarity tasks. The semantic closeness between the generated and reference summaries is then assessed using cosine similarity (SentenceBERT\_CS) and Euclidean distance (SentenceBERT\_ED).

\textbf{CodeT5\_plus\_CS}~\cite{wang2021codet5} uses CodeT5+ embeddings to compute cosine similarity between source code and summaries, enabling reference-free evaluation for code-aware contexts. 

\textbf{GPT\_emb} computes the cosine similarity between GPT-based embeddings of generated and reference summaries, capturing semantic relationships beyond n-gram overlap through high-dimensional contextual representations.



\textbf{TF-IDF}~\cite{ramos2003using} is a widely adopted term-weighting scheme that evaluates the significance of a word within a collection of documents and is typically used in combination with a distance metric, such as cosine similarity or Euclidean Distance.

Along the lines of embedding-based metrics, the \textbf{USE} (Universal Sentence Encoder)~\cite{cer2018universal} computes cosine similarity between encoded sentence vectors and is known for its scalability and generalization across tasks. 



\textbf{InferSent}~\cite{conneau2017supervised} uses GloVe embeddings~\cite{pennington2014glove} as pretrained word representations, processed through RNN-based encoders to yield fixed-length sentence vectors. As with other embedding-based metrics, cosine similarity and/or Euclidean distance are used to compare generated and reference summaries. 

While such methods often better align with human judgment~\cite{mastropaolo2024evaluating}, they still depend on reference summaries and may inherit their biases or inaccuracies. Moreover, as noted in \secref{sec:introduction}, Steidl \etal~\cite{Steidl:icpc2013} observed that developers frequently neglect to update comments after code changes--leading to increased risk of misalignment and reduced reliability in reference-based evaluation.


To address the limitations of generic evaluation metrics, Mastropaolo \etal~\cite{mastropaolo2024evaluating} proposed \textbf{SIDE}, a task-specific metric that captures structural and semantic alignment between code and summaries. Unlike traditional metrics, SIDE shows a stronger correlation with human judgments~\cite{roy2021reassessing} and offers a more reliable evaluation framework. Notably, SIDE is the first and only metric capable of evaluating summarization outputs without requiring reference summaries--making it particularly valuable in settings where human-written comments are unavailable.

However, SIDE was originally trained on Java and does not generalize to other languages without retraining. Despite this, we observed that some research has applied SIDE ``out-of-the-box'' to non-Java datasets---particularly Python---without adaptation~\cite{sun2024source}. This practice introduces serious semantic and methodological risks: SIDE's contrastive representations are fine-tuned to Java's syntax, naming conventions, and idioms, and may fail to meaningfully align code and summaries in other languages. 

To empirically validate this concern, we evaluated the off-the-shelf \side on the benchmark from Crupi \etal~\cite{crupi2025effectiveness}, which contains manually annotated Python $\langle$code, comment$\rangle$ pairs, achieving an average SIDE score of 0.65. 

To preserve SIDE's validity for Python, we retrained the contrastive model using the statistical framework and validation process described by Mastropaolo \etal~\cite{mastropaolo2024evaluating}. Our Python-specific variant---\sidep---achieves a substantially higher SIDE score of 0.90 on the same benchmark, demonstrating a 36.5\% improvement in alignment with human judgments. This significant performance gap empirically validates that language-specific retraining is essential for maintaining SIDE's effectiveness across different programming languages and preserving its goal of measuring semantic alignment in code summarization tasks.
Moreover, by replicating the original experimental setup and releasing our implementation publicly available on GitHub~\cite{replication}, 


While semantic evaluation metrics serve as essential tools for assessing summarization quality, they can also function as filters for identifying high-quality training data, enabling more efficient model training for code summarization. Building on this insight, recent work has shifted focus toward improving the training data itself rather than solely pursuing larger models or architectural innovations. 

\subsection{Optimizing Code Summarization Training Dataset via Coherence-Based Filtering}

While most work in code summarization has focused on model architecture and evaluation, limited attention has been paid to the quality of training data. Addressing this gap, Vitale \etal~\cite{vitale2025optimizing} proposed a data-centric approach grounded in the principle that ``not all training examples contribute equally to model learning.'' To operationalize this, they leveraged the SIDE metric to quantify semantic coherence between code and summaries, ranking and filtering out the least aligned 50\% of training pairs.

This alignment-based filtering yielded a smaller but semantically consistent dataset, representing a shift from the prevailing ``more data is better'' assumption toward a quality-over-quantity paradigm. Experiments with CodeT5+~\cite{wang2023codet5p} showed that, despite removing half of the training pairs, performance remained stable across BLEU, METEOR, and CodeBLEU scores, while models converged faster and required fewer epochs to reach peak performance. They also exhibited greater stability across random seeds, suggesting that alignment-based pruning improves not only efficiency but also robustness and reproducibility. Importantly, the study highlighted that models fine-tuned with SIDE-filtered data sometimes even surpassed the performance of those trained on the full datasets, particularly at moderate thresholds (\eg SIDE 0.6), showing that carefully pruning incoherent examples can sharpen the learning signal.

Vitale \etal further demonstrated that SIDE filtering can be applied incrementally: progressively discarding the least coherent examples sharpened the learning signal and produced efficiency gains without sudden performance collapse. This scalability enables the approach to be adaptable to various data budgets. By contrast, random pruning degraded accuracy, underscoring that dataset composition–not raw size–governs summarization quality.

Finally, Vitale \etal cautioned that training on semantically misaligned pairs does not merely waste computation but biases models toward superficial lexical overlap, encouraging memorization over functional understanding. They also noted that the persistence of strong results even under random pruning suggests that current benchmarks contain substantial redundancy, reinforcing the need to explore complementary quality attributes---such as readability and diversity---when curating code summarization datasets.

\subsection{Token-Based Data Optimization}

Complementing row-level filtering methods like SIDE, token-based optimization targets the input sequence itself, aiming to reduce verbosity at a finer granularity. Shi \etal~\cite{shi2022we} explored this direction by identifying low-utility tokens---such as rare terms, syntactic delimiters, or type annotations---that may contribute little or even hinder learning.

Through systematic ablation studies, they showed that moderate token reduction can retain, or even enhance, model performance. These findings support the hypothesis that essential summary-relevant information is concentrated in a small subset of tokens, and that reducing structural noise can improve both efficiency and output quality.

\section{Optimizing Code Summarization Training Dataset via Token-Level Optimization Techniques}

Motivated by recent calls for data-centric optimization in code-related tasks~\cite{shi2022we,yang2024less,vitale2025optimizing}, we introduce three reduction methods grounded in fine-grained token editing and domain-specific transformations. These methods are guided by the hypothesis that \textit{semantically irrelevant tokens can be removed without impairing the model's ability to generate meaningful summaries}. In practice, filtering out such tokens preserves core semantics while reducing redundancy and noise. This principle applies both to deliberate pruning---where low-utility tokens are explicitly targeted---and to structural transformations that implicitly eliminate them. For instance, converting code into its Abstract Syntax Tree (AST) representation naturally discards superficial syntactic tokens (\eg delimiters, formatting artifacts) while retaining essential structure.

To operationalize this hypothesis, we introduce:

\textbf{Abstract Syntax Tree (AST)} provides a structured, tree-based representation of source code based on the syntax and grammar of the programming language~\cite{alon2018code2seq}. Leaf nodes (terminals) denote low-level elements such as variable names and types, while non-terminals represent higher-level constructs including loops, expressions, and declarations. Due to its conciseness and structural fidelity, the AST is widely used as a proxy for raw code in downstream tasks~\cite{shi2020pathpair2vec, aladics2022ast}. In this work, we utilize ASTs as input representations for code summarization, replacing the raw source code.

\textbf{Function Signature} refers to the definition of a function, including its name, return type, and input parameters. In a study by Ding \etal~\cite{ding2024code}, the authors investigated four different input configurations for training code summarization models: (i) the full code sequence, (ii) the function signature alone, (iii) the function body (excluding the signature), and (iv) a modified function body where tokens identical to those in the signature are replaced with placeholder tokens (\eg \texttt{METHOD\_NAME}, \texttt{RETURN\_TYPE}, \texttt{VARIABLE\_TYPE}, \texttt{VARIABLE\_NAME}). Their findings indicate that the function signature alone---the header of the code snippet---is sufficient to yield strong performance in code summarization tasks. This result highlights the importance of the semantic and structural information encoded in function headers, suggesting that comment generation models should place greater emphasis on mining key signals from the function signature.

\textbf{CrystalBLEU}~\cite{eghbali2022crystalbleu} is a widely recognized evaluation metric designed to improve distinguishability in code summarization by discounting trivially shared common n-gram tokens that appear frequently across examples and contribute little to semantic differentiation. The core idea behind the metric is straightforward: first, identify the set of trivially shared n-grams, and then compute a modified BLEU score that excludes their influence. Inspired by this principle, our token reduction strategy utilizes CrystalBLEU to identify and eliminate trivially shared n-gram tokens from the input data, thereby reducing the overall training data volume while preserving semantically influential content for training.

\noindent The rationale for prioritizing these strategies follows four guiding pillars: 

1. \textbf{Explainability:} Each method is grounded in transparent, open-box rules for token reduction, ensuring that the rationale for every transformation is interpretable. For instance, CrystalBLEU identifies trivially frequent n-grams (\eg common keywords like \texttt{public} or \texttt{return}) and prunes them because they contribute little to distinguishing semantics across examples. Similarly, the elimination of syntactic delimiters or boilerplate tokens follows well-understood principles, making the outcomes of these transformations easy to analyze, reproduce, and trust.  

2. \textbf{Applicability Across Contexts:} The proposed techniques are computationally efficient, language-agnostic, and avoid the substantial overhead of approaches that rely on specialized pre-processing pipelines--such as graph-based representations that require graph encoders~\cite{leclair:icpc2020} or message-passing networks~\cite{zhou2020graph}. Such approaches often require costly conversions and custom neural architectures, which limit scalability and portability. In contrast, our strategies reduce verbosity through lightweight token- and structure-level transformations. Unlike black-box techniques such as LLM-driven token selection, where the basis of performance is opaque, our reductions follow explicit rules, making them both transferable across languages and straightforward to interpret.  


3. \textbf{Complementarity:} By spanning token-level filtering (\ie CrystalBLEU guided reduction), structural summarization (\ie AST-based transformation), and semantic restructuring (\ie function signature), the proposed methods capture distinct yet complementary dimensions of code information. CrystalBLEU focuses on statistical redundancy removal, AST preserves syntactic structure, and Function Signature extracts functional semantics---together providing a multi-faceted view of code representation that balances lexical, structural, and semantic properties.

To rigorously assess the impact of these reductions, we rely on Shannon entropy ~\cite{shannon1948mathematical} rather than a raw token count. Simply shortening sequences does not necessarily make them more efficient or more informative: two inputs of identical length can differ greatly in the amount of information they encode. \textbf{Shannon entropy}--a measure of the average information or uncertainty contained in a distribution of tokens--captures this nuance by quantifying how much meaningful content is carried by the retained sequence.

\begin{figure}[htbp]
\centering
\includegraphics[width=0.6\columnwidth]{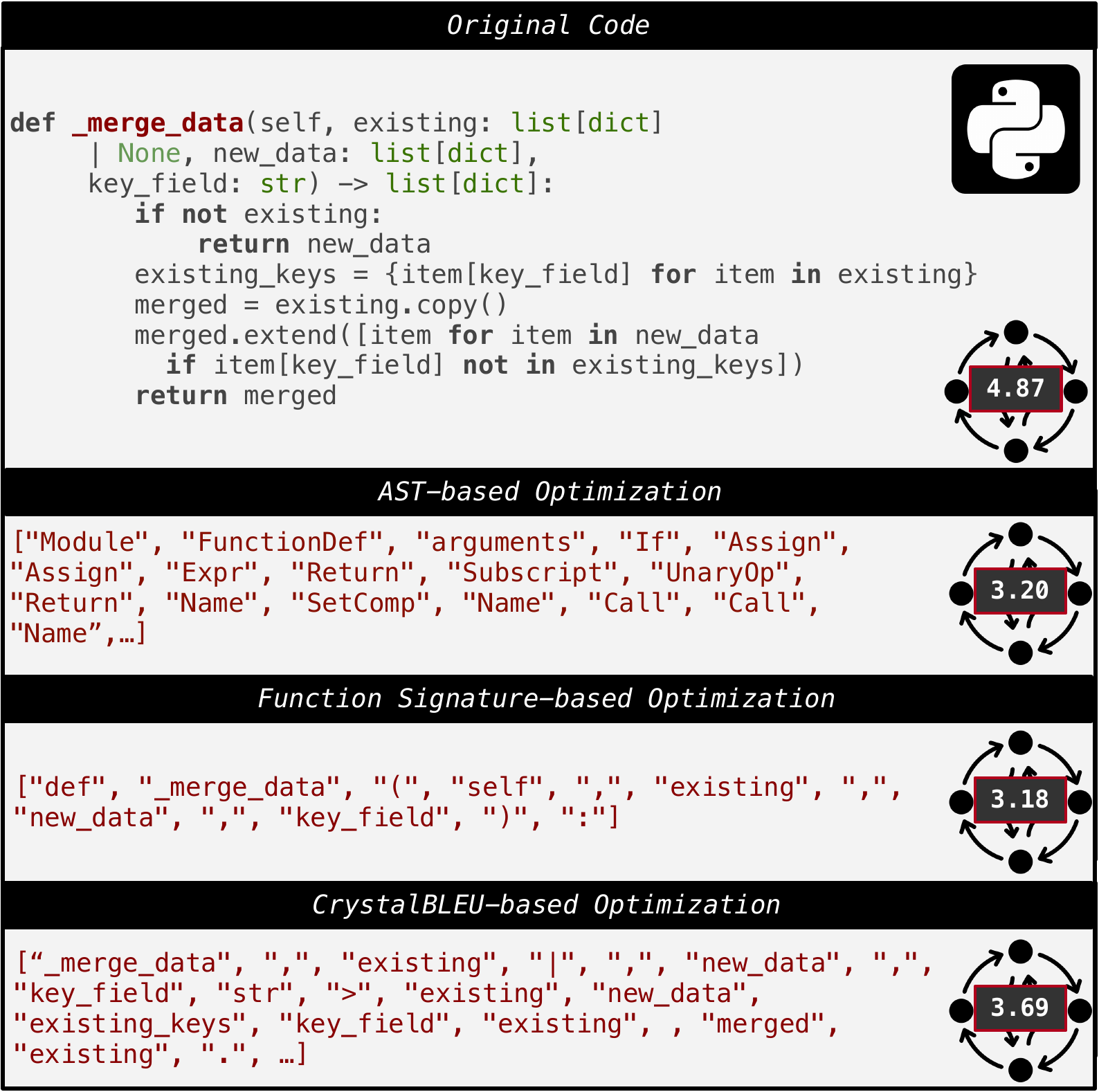}
\caption{Different Token Representation Techniques in Token-based Optimization and their computed Shannon entropy value}
\label{fig:token-rep}
\end{figure}

As shown in Figure~\ref{fig:token-rep}, the original method has an entropy of 4.87, whereas the optimized representations yield substantially lower values: 3.20 for the AST-based method, 3.18 for function signatures, and 3.69 when the shortening is CrystalBLEU-guided. These reductions demonstrate that the optimized inputs are not only shorter but also convey information more compactly and predictably.

This distinction is central to our setting: in code summarization, the goal is not just to reduce sequence length but to retain those tokens most critical for guiding the model toward concise and accurate summaries. Entropy reduction, therefore, serves as a principled indicator of effective compression, where redundant or boilerplate elements are removed while semantically meaningful cues are preserved. Empirical evidence supports this interpretation: models trained on entropy-optimized inputs converge faster, require fewer computational resources, and produce summaries that maintain--or sometimes even improve--overall informativeness, as corroborated by presented qualitative examples (\secref{sec:qa-implications}).

We test our hypothesis across these strategies and find consistent empirical support that token-level reductions not only yield substantial efficiency gains but, in some cases, improve summarization performance. Models trained on optimized inputs converge faster, consume fewer resources, and retain or even enhance informativeness

\vspace{-10pt}
\setlength{\textfloatsep}{10pt}
\section{Study overview}

To examine the impact of data-centric optimization on code summarization, we design three complementary studies:

\textbf{Study-1 (\secref{sec:study-1})}: We apply three token-level optimization strategies--AST-based optimization, function signature-based extraction(optimization), and CrystalBLEU-guided optimization--directly to Java code. Each technique is evaluated individually and in conjunction with the semantic filtering method by Vitale \etal~\cite{vitale2025optimizing}. This design supports a comprehensive analysis of isolated and layered reductions, revealing how syntactic and semantic optimizations may jointly impact data quality.

\textbf{Study-2 (\secref{sec:study-2})}: We develop and evaluate \sidep, a Python-adapted version of the SIDE (\textit{Summary alIgnment to coDe sEmantics}) metric introduced by Mastropaolo \etal~\cite{mastropaolo2024towards}. By replicating the original training pipeline with Python programs, we assess \sidep's ability to capture semantic alignment between code and summaries, as well as its correlation with human judgment, compared to standard metrics such as BLEU and ROUGE.

\begin{table*}[t]
\centering
\caption{Study overview (methodology only). Each study specifies its design, hypothesis, evaluation criteria, and main controls; results in Sections~5–7.}
\label{tab:study-overview}
\scriptsize
\renewcommand{\arraystretch}{1.0}
\begin{tabular}{p{0.15\textwidth} p{0.30\textwidth} p{0.47\textwidth}}
\toprule
\textbf{Study} & \textbf{Design \& Setup} & \textbf{Hypothesis \& Evaluation} \\
\midrule
\textbf{1: Token-level optimization (Java)} &
Three token reductions (AST, Signature, CrystalBLEU): (i) standalone on Funcom\textsubscript{Java}, (ii) cascaded on SIDE-filtered Funcom\textsubscript{Java}. Model: CodeT5+-220M, fixed training; metrics: BLEU, ROUGE-L, METEOR, token retention, entropy. &
\textbf{Hyp.:} Token reduction lowers compute while preserving quality. \newline
\textbf{Eval.:} Token retention (\%), entropy, automatic metrics. 
\\
\midrule
\textbf{2: \(\text{SIDE}_{\text{py}}\) (Python)} &
Retrain SIDE contrastive framework for Python using original protocol; evaluate via human judgments (Crupi \etal) and lexical/semantic metrics. &
\textbf{Hyp.:} \(\text{SIDE}_{\text{py}}\) aligns better with human adequacy than lexical baselines. \newline
\textbf{Eval.:} Human correlation, logistic regression ORs.
\\
\midrule
\textbf{3: Replication for Generalizability (Python)} &
Replicate Study~1's reductions on Funcom\textsubscript{Python}, varying only language. Cascade with \(\text{SIDE}_{\text{py}}\); evaluate on new 500-item human-annotated Python benchmark + CoderEval-Python; measure token retention (\%) and entropy. &
\textbf{Hyp.:} Study~1's trade-offs generalize across languages. \newline
\textbf{Eval.:} Token retention (\%), entropy, automatic metrics, cross-language consistency.
\\
\bottomrule
\end{tabular}
\end{table*}

\textbf{Study-3 (\secref{sec:study-3})}: We extend Study-1 to Python to assess the language-agnostic nature of the token-based optimization methods and examine their generalizability. In this setting, we apply \sidep to filter misaligned \ccpair pairs, combining semantic filtering~\cite{vitale2025optimizing} with the token-level strategies used in Study-1. This study investigates whether the methods remain effective across different programming languages.

Table~\ref{tab:study-overview} summarizes the methodological design of our three studies, highlighting their setup, guiding hypotheses, evaluation criteria, and controls. Together, these studies provide a coherent framework: Study~1 establishes the role of token-level reductions in Java, Study~2 extends semantic alignment evaluation to Python through \(\text{SIDE}_{\text{py}}\), and Study~3 tests the cross-language robustness of both approaches. The following sections (Sections~5–7) detail the implementation and evaluation of each study in turn.

Methodological and evaluation details for each study are presented in the following sections.

\section{Study 1: Optimizing Code Summarization Training Dataset via Token-based Reduction Methods}

\label{sec:study-1}

\textit{\textbf{Study 1}} investigates token-level optimization techniques to enhance training efficiency for code summarization, aiming to reduce training time and computational costs. Building on the semantic filtering approach of Vitale \etal~\cite{vitale2025optimizing} through the \side metric, we investigate whether additional reductions can be achieved without degrading performance. 

To this end, each token-level strategy is evaluated on both the unoptimized dataset and the SIDE-filtered dataset, enabling a direct comparison between standalone and cascading configurations. The latter simply applies token-level reductions after semantic filtering, allowing us to assess whether combining the two stages yields complementary benefits or introduces trade-offs compared to using either strategy alone.

\noindent This study addresses the following research questions: 

\vspace{5pt}

\noindent \textbf{\rq{1}: \textit{How effective are token-based optimization techniques for code summarization compared to established state-of-the-art alignment methods, such as \side?}}. We investigate the standalone impact of applying token-based optimization without alignment filtering (as provided by \side). This analysis aims to establish a baseline performance that is independent of any cascading effects introduced by combining multiple strategies.

\noindent \textbf{\rq{2}: \textit{What is the impact of cascading optimization strategies on the performance of code summarization models?}}
We examine the effectiveness of combining token-based optimization techniques with semantic alignment-based filtering in a cascading pipeline. The goal is to determine whether cascaded optimization yields complementary improvements to the semantic alignment-based filtering compared to applying each approach in isolation, or instead introduces trade-offs in the performance of code summarization models.

\subsection{Dataset Optimization}
\label{subsec:data-op}
To address \textbf{RQ\textsubscript{1}}, we leverage the Funcom dataset \cite{leclair:icse2019}, a widely used benchmark specifically curated for code summarization tasks \cite{leclair:icse2019}. Funcom originally comprised over 2.1 million \ccpair pairs extracted from the Sourcerer repository \cite{lopes2010sourcerer}. However, as noted by Shi \etal~\cite{shi2022we}--the seminal dataset presented by LeClaire \etal suffered from substantial overlap, noise, and duplicate entries. To mitigate these issues, Shi \etal apply a heuristic-based cleaning methodology. We adopt their cleaned version of Funcom, which contains 1,184,438 training instances, as the foundation for evaluating the effects of our optimization techniques.
Additionally, we adapt the \textit{SIDE-filtered Funcom} data from the study by Vitale \etal, which applies semantic alignment filtering with varying levels of coherence-retention (\eg \textsc{SIDE}$\geq0.7$, \textsc{SIDE}$\geq0.8$ \etc). 
For our experiments, we specifically opt for the semantic-filter cutoff at \textsc{SIDE}$\geq0.9$ as it achieves the largest dataset reduction ($\approx 51$--$54\%$) while maintaining performance statistically indistinguishable from the full data \cite{vitale2025optimizing}. This threshold represents the most cost-effective point on the Pareto frontier, is used for the paper's Random baseline comparison, and yields substantial efficiency gains (\eg $\sim111$\,h less training time and $-55\%$ CO$_2$ on Funcom\textsubscript{Java}). Thus, it provides the highest-coherence stratum for our cascading optimization (semantic pruning $\rightarrow$ token-level compaction) without sacrificing quality.

To implement the \textbf{\astApproach} method, we rely on the \texttt{JavaLang} parser~\cite{c2nes2020javalang}, a widely used tool for converting Java source code into its AST representation. 
This parser has been employed in prior research~\cite{aladics2022ast,shi2020pathpair2vec}, providing empirical evidence of its use for extracting syntactic and structural information from code. 
In our setting, these structural elements are projected into a sequence of tokens forming a well-defined AST representation~\cite{alon2018code2seq, mastropaolo2021empirical}. 
Therefore, \texttt{JavaLang}  provides a consistent mechanism to transform every method into an AST-based token sequence, which we can directly substitute for raw lexical tokens in our experiments. 
For each method in the dataset, we parse the source code with \texttt{JavaLang} and extract the corresponding AST tokens. 
These tokens provide a structured abstraction of the original code and replace raw code tokens as the model input.
This AST-based transformation reduces the token count by 55.92\% on the cleaned Funcom dataset---bringing it down from 100\% to 44.08\% of the original. In the visualizations that follow, the icon \reductionicon{44.08} represents bars divided into two sections: the purple segment corresponds to the original token count (normalized to 100\%), while the green segment shows the remaining tokens after applying one of the three optimization strategies. In this case, 44.08\% of the tokens remain after transformation (100\% − 55.92\% = 44.08\%), which is the value illustrated by the green bar.

The second strategy, the \textbf{\functionApproach}, is inspired by Ding \etal~\cite{ding2024code}. For each code instance, we extracted only the function signature--consisting of the method name, return type, and parameter list--using the same \texttt{JavaLang} parser. This design focuses on the semantic and structural cues developers rely on to quickly understand code. The resulting signature tokens replaced the full method body as the model input. When measuring token reduction, this technique achieves a 69.51\% \reductionicon{30.49} reduction compared to the full code representation.

%

Finally, the third strategy, the \textbf{\cbleuApproach}~\cite{eghbali2022crystalbleu}, adapts CrystalBLEU from an evaluation metric into a reduction mechanism. 
In this approach, we identify and discard \textit{trivially shared $n$-grams}--that is, frequent patterns that appear across many examples but provide little semantic discrimination (\eg common keywords such as \texttt{public} or \texttt{return}, delimiters, or boilerplate constructs). This process retains only tokens likely to contribute to meaningful differentiation across examples. 
On the cleaned Funcom\textsubscript{Java} dataset, this method produced a 72.23\% \reductionicon{27.77} reduction in token count compared to the original code.



\subsection{Model Training}
\label{subsec:model-training}

To answer \rq{1} and \rq{2}, we adopt a pre-trained Transformer model CodeT5+~\cite{wang2023codet5p} as our backbone model, given its demonstrated effectiveness across a range of code-related tasks~\cite{mastropaolo2024towards, ahmed2024automatic, phan2024repohyper}, including summarization. As an extension of the T5 architecture~\cite{raffel2020exploring}, CodeT5+ incorporates additional objectives for source code modeling, including span denoising, causal language modeling~\cite{tay2022ul2, soltan2022alexatm}, and text-code contrastive learning.

Among CodeT5+ available sizes (220M–16B), we select the 220M variant to balance efficiency and representational capacity. All models are fine-tuned for 20 epochs with a batch size of 16, a maximum input length of 512 tokens, and output capped at 128 tokens--following prior setups on Funcom~\cite{mastropaolo2022using, zhou2022automatic, vitale2025optimizing}. We use the AdamW optimizer~\cite{loshchilov2017decoupled} with a learning rate of 2e-5~\cite{ciniselli2024generalizability, mastropaolo2024training}, and apply early stopping with patience of five epochs and a minimum delta of 0.01~\cite{mastropaolo2023towards, Ciniselli:icpc2023}. 


We fine-tune CodeT5+ on eight versions of the Funcom dataset to systematically evaluate token optimization strategies: four configurations apply our three token-level reductions (AST, Function Signature, CrystalBLEU) plus an (un)-optimized baseline to the original cleaned Funcom data, and four parallel configurations apply the same optimizations to the SIDE-filtered variant from Vitale \etal~\cite{vitale2025optimizing}. This design enables direct comparison of standalone token optimization effects and cascaded optimization, where token reduction is applied after semantic filtering.

Performance is monitored on the validation set after each epoch, and the best checkpoint is retained if no further improvements are observed after 5 consecutive epochs (patience) with a minimum delta of 0.01. This procedure is independently repeated for each of the eight optimized datasets described in \secref{subsec:data-op}. Table \ref{tab:training-config} summarizes the training configuration, including hyperparameter settings.

\begin{table}[h]
\centering
\caption{Model training configuration for CodeT5+ (220M) fine-tuning on Funcom datasets}
\label{tab:training-config}
\small
\begin{tabular}{lll}
\toprule
\textbf{Configuration} & \textbf{Parameter} & \textbf{Value/Details} \\
\midrule
{\textbf{Model Architecture}} 
& Base Model & CodeT5+ (220M parameters) \\
\midrule
\multirow{6}{*}{\textbf{Training Hyperparameters}} 
& Epochs & 20 \\
& Batch Size & 16 \\
& Optimizer & AdamW \\
& Learning Rate & 2e-5 \\
& Max Input Length & 512 tokens \\
& Max Output Length & 128 tokens \\
& Early Stopping & Patience: 5 epochs, min delta: 0.01 \\
\midrule
\multirow{10}{*}{\textbf{Dataset Configurations}} 
& \multicolumn{2}{l}{\textbf{Total Models Trained: 16 (8 Java + 8 Python)}} \\
\cmidrule(l){2-3}
& \multicolumn{2}{l}{\textit{Java (8 configurations):}} \\
& \quad Standalone & Original: Baseline, AST, Signature, CrystalBLEU \\
& \quad SIDE Optimized & SIDE-filtered: Baseline, AST, Signature, CrystalBLEU \\
\cmidrule(l){2-3}
& \multicolumn{2}{l}{\textit{Python (8 configurations):}} \\
& \quad Standalone & Original: Baseline, AST, Signature, CrystalBLEU \\
& \quad SIDE Optimized & SIDE-filtered: Baseline, AST, Signature, CrystalBLEU \\
\cmidrule(l){2-3}
& Evaluation Metrics & BLEU, ROUGE-L, METEOR, token retention, entropy \\
\bottomrule
\end{tabular}
\end{table}

\subsection{Experiments and Evaluation}
\subsubsection{\textbf{Evaluation Dataset:}}

Mimicking the evaluation of Vitale \etal, we employ the CoderEval dataset to assess our fine-tuned models, which has been specifically adapted for code summarization tasks. CoderEval \cite{yu2024codereval} was originally designed for code generation and comprises 230 manually curated Java problems drawn from high-quality, highly starred open-source repositories. Each instance pairs a human-written docstring---intended as a generation prompt---with its corresponding code implementation. To repurpose the dataset for summarization, Vitale \etal inverted the original $\langle \textit{docstring}, \textit{code} \rangle$ format into $\langle \textit{code}, \textit{docstring} \rangle$, enabling evaluation of models that produce natural-language summaries from source code. A subsequent manual review ensured that all examples reflected realistic development scenarios and maintained semantic alignment between code and comments. The final Java subset used in our experiments contains 218 high-quality $\langle \textit{code}, \textit{summary} \rangle$ pairs.

Following Vitale \etal, to strengthen our evaluation, we complement our assessment with the dataset curated by Mastropaolo \etal~\cite{Mastropaolo:icse2023}, which includes 892 Java methods paired with summaries corresponding to the first sentence of their documentation.
This dataset was built from non-fork GitHub repositories that satisfied stringent quality thresholds ($\geq 300$ commits, $\geq 50$ contributors, and $\geq 25$ stars), and underwent systematic quality checks to ensure reliability. Vitale \etal later refined and extended this resource, resulting in a dataset subjected to multiple rounds of validation and refinement, which we leverage for robust cross-evaluation of our token optimization strategies.

\subsubsection{\textbf{Evaluation Metrics:}}

To quantify the trade-off between input reduction and downstream performance in optimized training datasets for code summarization, we adopt a standard set of metrics consistent with those used by Vitale \etal~\cite{vitale2025optimizing}.
We note that these represent only a subset of the metrics (BLEU, METEOR, ROUGE) used in our second study (see \secref{sec:study-2}). A comprehensive overview of the evaluation metrics can be found in \secref{sec:related}.


Shannon entropy reduction quantifies the decrease in information-theoretic complexity of the training data, reflecting how effectively each method minimizes redundancy while retaining essential information.
To compute this measure, we tokenized the entire dataset using the model's tokenizer and calculated the Shannon entropy over the resulting token distribution. Specifically, for each representation (original and optimized), we compute 
\begin{equation}
    H = -\sum_{i} p_i \log_2 p_i
\end{equation}
where $p_i$ is the normalized frequency of token i in the dataset. The entropy reduction was then obtained by comparing the entropy of the optimized representation against that of the original (baseline) representation.

This approach allows us to assess how much each optimization compresses the informational content of the dataset while preserving meaningful variation relevant to the model’s learning process.

We do not report floating-point operations (FLOPs) in this study, as the total number of training instances remains constant across conditions (as in our experimental design), FLOPs measurements become identical regardless of input token variations, making them uninformative for comparing token-level optimizations. 
\subsubsection{\textbf{Results:}}

The results in \tabref{tab:res-1_1} and \tabref{tab:res-1_2} illustrate the trade-offs between token-level reductions and summarization performance across two evaluation datasets, with Shannon entropy serving as our primary measure of information compression effectiveness.

Examining the \textbf{standalone token-optimization} results in \tabref{tab:res-1_1}, the AST-based approach clearly stands out on the CoderEval dataset, achieving a BLEU score of 8.67 while retaining only 44.08\% of tokens \reductionicon{44.08}, a remarkable 37\% improvement over the baseline score of 6.32. Considering the additional metrics, ROUGE-L (38.82) and METEOR (32.24), we can conclude that structural information alone is sufficient to produce code summaries that are on par with, and in some cases even superior to, those generated by more complex code representations.

For code summarization tasks, optimizing inputs using CrystalBLEU achieves the most substantial compression, reducing entropy by 33.20\% while preserving only 27.65\% \reductionicon{27.65} of the original tokens. As information is progressively removed from the input methods, we observe a proportional decline in performance, reflected in BLEU, ROUGE-L, and METEOR scores, which show losses ranging from approximately $\sim$1.5\% to $\sim$7\% when evaluated on CoderEval. Interestingly, \cbleuApproach unexpectedly delivers the best results when tested against Mastropaolo \etal's benchmark, indicating the importance of input data characteristics. 
To understand this surprising behavior, we analyzed the average token counts across both datasets. Mastropaolo \etal's methods average 167.77 tokens compared to CoderEval's 69.00 tokens--a 2.4-fold increase in code complexity. After CrystalBLEU optimization, Mastropaolo \etal's methods reduce to 55.42 tokens while CoderEval reduces to 18.07 tokens. This pattern reveals that CrystalBLEU's frequency-based pruning becomes more effective on verbose and complex code, where redundant patterns, such as common keywords, constitute a larger proportion of input tokens. By removing this noise from longer methods, CrystalBLEU enables models to focus on semantically meaningful patterns, which explains its superior performance on the more challenging Mastropaolo benchmark, despite the dataset's overall difficulty.


The Function Signature-based optimization retains only 30.49\% \reductionicon{30.49} of tokens while reducing entropy by 29.82\%. On CoderEval, BLEU decreases modestly from 6.32 (baseline) to 6.23--a minimal 1.4\% decline despite eliminating 69.51\% of input tokens. Similar patterns emerge for ROUGE-L and METEOR, which show proportionally small degradation relative to the substantial compression achieved. On the Mastropaolo \etal benchmark, this method maintains identical BLEU performance (1.88) to \cbleuApproach (the second-best technique) while showing slight reductions in ROUGE-L and METEOR scores.

Despite the expected performance trade-offs relative to the baseline, an intriguing cross-benchmark pattern emerges. Mastropaolo exhibits substantially lower baseline scores across all metrics (BLEU: 1.80 vs. 6.32 on CoderEval), indicating a considerably more challenging evaluation context---likely due to increased code complexity or reduced documentation quality. This observation aligns with our earlier statement regarding input complexity: Mastropaolo methods average 167.77 tokens, compared to CoderEval's 69.00 tokens, representing 2.4 times more structural and lexical information that models must process to generate comparably sized summaries (15.00 vs. 12.24 tokens on average). Therefore, it looks like under these demanding conditions, compression-based approaches demonstrate unexpected robustness and, in some cases, improvement. This counterintuitive finding suggests that strategic token reduction may actually enhance model performance on structurally complex code. By distilling inputs to their most semantically dense components, compression techniques appear to help the model filter noise and focus on core semantic patterns that drive accurate summarization.


Building on these standalone observations, we next analyze the \textbf{cascading effect} of combining semantic filtering with token-level reductions. The results in \tabref{tab:res-1_2} reveal a clear pattern: semantic pre-filtering fundamentally transforms the effectiveness of token optimization. On CoderEval, all three methods maintain the SIDE baseline BLEU of 6.32, demonstrating that aggressive token reduction--retaining only 27-44\% of tokens--can be applied without quality loss when operating on semantically coherent data. AST continues to lead, achieving 7.04 BLEU with the strongest ROUGE-L (36.28) and METEOR (30.06), representing an 11\% improvement over the SIDE baseline. This contrasts sharply with standalone results, where AST's advantage was even more pronounced (8.67 vs. 6.32). 

\begin{table*}[t]
\centering
\caption{Evaluation of \textbf{standalone token-optimization} strategies on Funcom\textsubscript{Java} using CodeT5+ (220M). The second column shows the entropy for each optimization technique, with parenthesized values indicating the percentage of entropy reduction. Best scores per metric are in \textbf{bold}.}
\label{tab:res-1_1}
\footnotesize
\resizebox{\textwidth}{!}{%
\begin{tabular}{lccccccccc}
\toprule
\multirow{3}{*}{\textbf{Approach}} &
\multirow{3}{*}{\textbf{\begin{tabular}[c]{@{}c@{}}Entropy Reduction\\ (vs. Original)\end{tabular}}} &
\multirow{3}{*}{\textbf{\begin{tabular}[c]{@{}c@{}}Token\\Retention\end{tabular}}} &
\multicolumn{3}{c}{\textbf{CoderEval \cite{yu2024codereval}}} &
\multicolumn{3}{c}{\textbf{Mastropaolo \etal \cite{Mastropaolo:icse2023}}}
\\
\cmidrule(lr){4-6}\cmidrule(lr){7-9}
& & & \textbf{BLEU} & \textbf{ROUGE-L} & \textbf{METEOR} & \textbf{BLEU} & \textbf{ROUGE-L} & \textbf{METEOR} \\
\midrule
Original Funcom\textsubscript{Java} & 4.095     & 100\% & \textbf{6.32} & 34.09 & 28.64 & 1.80 & 10.73 & 9.10 \\
\midrule
\cbleuApproach                      & \bf \cellcolor{yellow!60!orange!30}2.735 (\textcolor{red}{$\downarrow$}33.20\%)  & 27.65\% & 4.88 & 27.61 & 22.66 & 1.88 & \textbf{12.28} & \textbf{9.88} \\
\midrule
\functionApproach               & 2.874 (\textcolor{red}{$\downarrow$}29.82\%)  & 30.49\% & 6.23 & 31.39 & 26.07 & 1.88 & 11.64 & 9.53 \\
\midrule
\astApproach                             & 3.246 (\textcolor{red}{$\downarrow$}20.73\%)  & 44.08\% & \textbf{8.67} & \textbf{38.82} & \textbf{32.24} & 1.83 & 10.92 & 9.18 \\
\midrule
\rowcolor{blue!10} 
SIDE Optimized Funcom\textsubscript{Java} & 
\royalval{4.105} & 
\royalval{--} &
\royalval{6.32} & 
\royalval{34.15} & 
\royalval{27.72} & 
\royalval{1.94} & 
\royalval{11.82} & 
\royalval{10.00} \\
\bottomrule
\end{tabular}%
}
\end{table*}

\begin{table*}[t]
\centering
\caption{Evaluation of \textbf{token-optimization} strategies on the \textbf{SIDE Optimized} Funcom\textsubscript{Java} dataset using CodeT5+ (220M). The second column shows the entropy value for each optimization technique, with parenthesized values indicating the percentage of entropy reduction. Best scores per metric are in \textbf{bold}.}
\label{tab:res-1_2}
\footnotesize
\resizebox{\textwidth}{!}{%
\begin{tabular}{lccccccccc}
\toprule
\multirow{3}{*}{\textbf{Approach}} & 
\multirow{3}{*}{\textbf{\begin{tabular}[c]{@{}c@{}}Entropy Reduction\\ (vs. Original)\end{tabular}}} & 
\multirow{3}{*}{\textbf{\begin{tabular}[c]{@{}c@{}}Token\\Retention\end{tabular}}} &
\multicolumn{3}{c}{\textbf{CoderEval\cite{yu2024codereval}}} & 
\multicolumn{3}{c}{\textbf{Mastropaolo \etal\cite{Mastropaolo:icse2023}}} \\
\cmidrule(lr){4-6} \cmidrule(lr){7-9}
& & & \textbf{BLEU} & \textbf{ROUGE-L} & \textbf{METEOR} & \textbf{BLEU} & \textbf{ROUGE-L} & \textbf{METEOR} \\
\midrule
SIDE Optimized Funcom\textsubscript{Java} & 4.105 & 100\% & \textbf{6.32} & 34.15 & 27.72 & \textbf{1.94} & \textbf{11.82} & \textbf{10.00} \\
\midrule
\cbleuApproach & \bf \cellcolor{yellow!60!orange!30}2.736 (\textcolor{red}{$\downarrow$}33.35\%) & 27.77\% & 6.32 & 31.66 & 26.33 & 1.80 & 10.93 & 9.21 \\
\midrule
\functionApproach & 2.862 (\textcolor{red}{$\downarrow$}30.17\%) & 28.65\% & 5.94 & 32.37 & 26.92 & 1.89 & 11.47 & 9.61 \\
\midrule
\astApproach & 3.264 (\textcolor{red}{$\downarrow$}20.48\%) & 44.31\% & \textbf{7.04} & \textbf{36.28} & \textbf{30.06} & 1.87 & 10.73 & 9.38 \\
\bottomrule
\end{tabular}%
}
\end{table*}
\begin{table*}[t]
\centering
\caption{Statistical comparison of approaches for Java using SIDE-Optimized and Standalone Token-Optimized datasets. P-values are reported as ``x'' when $\geq$ 0.05 (not statistically significant) and as the actual value when $< 0.05$ (significant). Cliff's Delta effect sizes are provided for all comparisons with labels indicating magnitude: N (Negligible: $|\delta| < 0.147$), S (Small: $0.147 \leq |\delta| < 0.33$), M (Medium: $0.33 \leq |\delta| < 0.474$), L (Large: $|\delta| \geq 0.474$).}
\label{tab:statistical-st1}
\footnotesize
\resizebox{\textwidth}{!}{%
\begin{tabular}{llcccccccccccc}
\toprule
\multirow{3}{*}{\textbf{Dataset}} & \multirow{3}{*}{\textbf{Metrics}} & \multicolumn{6}{c}{\textbf{CoderEval \cite{yu2024codereval}}} & \multicolumn{6}{c}{\textbf{Mastropaolo \etal \cite{Mastropaolo:icse2023}}} \\
\cmidrule(lr){3-8} \cmidrule(lr){9-14}
& & \multicolumn{2}{c}{\textbf{CrystalBLEU}} & \multicolumn{2}{c}{\textbf{Function Signature}} & \multicolumn{2}{c}{\textbf{AST}} & \multicolumn{2}{c}{\textbf{CrystalBLEU}} & \multicolumn{2}{c}{\textbf{Function Signature}} & \multicolumn{2}{c}{\textbf{AST}} \\
\cmidrule(lr){3-4} \cmidrule(lr){5-6} \cmidrule(lr){7-8} \cmidrule(lr){9-10} \cmidrule(lr){11-12} \cmidrule(lr){13-14}
& & \textbf{p-value} & \textbf{Cliff's $\delta$} & \textbf{p-value} & \textbf{Cliff's $\delta$} & \textbf{p-value} & \textbf{Cliff's $\delta$} & \textbf{p-value} & \textbf{Cliff's $\delta$} & \textbf{p-value} & \textbf{Cliff's $\delta$} & \textbf{p-value} & \textbf{Cliff's $\delta$} \\
\midrule
\multirow{3}{*}{SIDE Optimized Funcom\textsubscript{Java}} 
& BLEU & x & -0.048 (N) & x & -0.072 (N) & x & 0.047 (N) & <0.001 & -0.075 (N) & x & -0.011 (N) & <0.005 & -0.059 (N) \\
& ROUGE-L & x & -0.084 (N) & x & -0.053 (N) & 0.037 & 0.081 (N) & 0.001 & -0.067 (N) & x & -0.028 (N) & <0.001 & -0.079 (N) \\
& METEOR & x & -0.070 (N) & x & -0.035 (N) & x & 0.070 (N) & <0.001 & -0.089 (N) & x & -0.041 (N) & <0.001 & -0.087 (N) \\
\midrule
\multirow{3}{*}{Standalone Token-Optimized Funcom\textsubscript{Java}} 
& BLEU & <0.001 & -0.182 (S) & x & -0.032 (N) & 0.003 & 0.105 (N) & x & 0.020 (N) & 0.012 & 0.052 (N) & x & 0.012 (N) \\
& ROUGE-L & <0.001 & -0.210 (S) & 0.035 & -0.100 (N) & <0.001 & 0.133 (N) & <0.001 & 0.093 (N) & 0.009 & 0.062 (N) & x & 0.010 (N) \\
& METEOR & <0.001 & -0.205 (S) & x & -0.091 (N) & 0.007 & 0.095 (N) & 0.003 & 0.063 (N) & 0.024 & 0.072 (N) & x & 0.020 (N) \\
\bottomrule
\end{tabular}%
}
\end{table*}


While CoderEval results align with patterns observed in standalone token optimization, the Mastropaolo \etal benchmark reveals a contrasting trend. Here, applying token-reduction methods after semantic filtering results in modest performance declines, with BLEU scores ranging from 1.80 to 1.89, compared to the 1.94 BLEU score of the SIDE baseline. This pattern becomes evident when examining the Mastropaolo \etal columns in \tabref{tab:res-1_2}, where compressed inputs following semantic filtering yield diminished returns compared to their CoderEval counterparts.
This finding reinforces that Mastropaolo's more challenging benchmark---characterized by greater code complexity (167.77 vs. 69.00 average tokens)---requires preserving richer lexical detail for accurate summarization. Conversely, CoderEval's shorter and more structured examples tolerate aggressive compression, as essential semantic information remains accessible despite substantial token reduction.

Finally, examining entropy reduction across semantically filtered data reveals a clear divergence: on CoderEval, AST's moderate compression (44.31\% \reductionicon{44.31} token retention) achieves the strongest performance (BLEU: 7.04), whereas on Mastropaolo's benchmark, optimal results are obtained without any token-reduction methods applied post-filtering. This outcome aligns with our earlier observations, underscoring that for complex, challenging code, preserving complete lexical and structural information is critical. Further compression beyond semantic filtering risks eliminating subtle cues necessary for accurate summarization.

To validate these findings, we performed Wilcoxon signed-rank tests with Cliff's delta effect size calculations, comparing each token optimization technique against its respective baseline through paired comparisons (\tabref{tab:statistical-st1}). For SIDE-optimized data, we compared each token method against the SIDE-filtered baseline; for standalone optimizations, we compared against the original unoptimized dataset. To account for multiple comparisons across three methods and three metrics, we applied Holm-Bonferroni correction \cite{holm1979} to adjust p-values and control the familywise error rate. The statistical analysis reveals nuanced patterns that vary by dataset and metric.

For the SIDE-optimized dataset on CoderEval, no token optimization method shows statistically significant differences (all p-values $>$ 0.05), with negligible effect sizes across all comparisons (Cliff's delta ranging from -0.048 to -0.084 for CrystalBLEU, -0.032 to -0.072 for Function Signature, and 0.047 to 0.081 for AST). This statistical equivalence, despite 56-72\% token reduction, demonstrates that semantic filtering enables aggressive compression without measurable quality degradation. On the Mastropaolo et al. benchmark, while AST shows statistically significant differences across all metrics (p $<$ 0.05), the effect sizes remain negligible (Cliff's delta: -0.059 to -0.087), indicating that although differences are statistically detectable, their practical magnitude is minimal. CrystalBLEU and Function Signature maintain non-significant differences with negligible effect sizes on this benchmark as well.

In contrast, the standalone token-optimized dataset reveals more pronounced effects. CrystalBLEU demonstrates statistically significant differences (p $<$ 0.05) across all metrics on CoderEval, with small negative effect sizes (Cliff's delta: -0.182 to -0.210), indicating measurable but modest performance degradation when applied to unfiltered data. Notably, AST shows negligible-to-small positive effects on CoderEval (Cliff's delta: 0.095 to 0.133 for METEOR and ROUGE-L), with ROUGE-L reaching statistical significance (p $<$ 0.001), confirming that structural abstraction can improve certain aspects of summary quality even without semantic pre-filtering. On the Mastropaolo et al. benchmark, all three methods exhibit negligible-to-small effect sizes (ranging from 0.010 to 0.093), with most comparisons achieving statistical significance. However, the practical impact remains modest.

These statistical patterns reinforce our primary findings: semantic filtering can serve as a foundation for aggressive token compression without compromising quality, and the type of tokens preserved (structural vs. lexical) is more important than the compression rate alone. Complete statistical results are available in our replication package~\cite{replication}.

\vspace{5pt}

\begin{tcolorbox}[
    colback=cyan!8, 
    colframe=black, 
    coltext=black,
    arc=6pt, 
    boxrule=0.8pt, 
    left=5pt, 
    right=5pt, 
    top=8pt, 
    bottom=8pt, 
    fonttitle=\bfseries,
    coltitle=black, 
    title=Study 1 Summary,
    enhanced,
    attach boxed title to top left={yshift=-3mm, xshift=5mm},
    boxed title style={
        colback=gray!40,
        boxrule=0.7pt,
        arc=8pt,
        outer arc=8pt,
        left=5pt,
        right=5pt,
        top=0.5pt,
        bottom=0.5pt,
    }
]

The obtained findings reveal key insights into token-level optimization. 
Semantic filtering enables aggressive token compression (56-72\% reduction) without quality degradation, whereas unoptimized data shows measurable declines under the same reductions. Each strategy operates through distinct mechanisms: CrystalBLEU removes statistical redundancy (resilient on difficult datasets), Function Signatures emphasize semantic cues (inconsistent results), and AST preserves syntactic information (strongest performance).
Last but not least, the choice of the optimal strategy depend on deployment priorities and dataset characteristics. 
\end{tcolorbox}

\vspace{-10pt}
\setlength{\textfloatsep}{10pt}

\section{Study 2: Extending  Semantic Alignment from Java to Python with \sidep}
\label{sec:study-2}
Study 2 develops a Python-specific variant of the semantic alignment–based evaluation metric \side--extending the previous contribution by Mastropaolo \etal \cite{mastropaolo2024evaluating}. In particular, the original SIDE metric \cite{mastropaolo2024evaluating} employs a contrastive learning framework to model semantic similarity between code snippets and their summaries, but its implementation was limited to Java $\langle code, comment\rangle$ pairs. Although SIDE has been applied to other programming languages without any retraining or proper validation, a practice that can undermine model reliability and domain alignment~\cite{sun2024source}, we extend the framework by replicating its full development pipeline for Python, thereby enabling language-specific semantic evaluation in Python-based code summarization tasks.

Thus, to evaluate the reliability and effectiveness of the \sidep metric, the study addresses the following research questions:

\textbf{RQ\textsubscript{1}: \textit{Does the \sidep metric reliably evaluate code summarization quality in line with human judgments?}}
This research question examines the validity of the \sidep metric as an automatic evaluation tool for code summarization. In particular, it seeks to determine whether \sidep can reliably differentiate between high- and low-quality summaries and produce evaluation scores that are strongly correlated with human judgments of adequacy and informativeness. The goal is to assess the extent to which \sidep can serve as a consistent and meaningful proxy for human evaluation in the context of code summarization tasks.

\textbf{RQ\textsubscript{2}: \textit{To what extent does \sidep correlate with human judgments of summary quality and with traditional evaluation metrics?}}
This research question examines the extent to which the Python-adapted \sidep metric aligns with human assessments of code summary quality, relative to widely used metrics such as BLEU, ROUGE, METEOR, and embedding-based approaches like BERTScore and CodeT5+. The objective is to evaluate \sidep's effectiveness as a semantic evaluation metric by comparing its correlation with human ratings of summary adequacy and informativeness, thereby assessing its potential as a more reliable proxy for human judgment in Python-based summarization tasks.

The following subsections present the methodology for constructing \sidep and outline the procedures used to evaluate its effectiveness as a semantic code summarization metric.

\subsection{Measuring Alignment between Python Methods and Docstrings with \sidep}
\subsubsection{\textbf{Finetuning Dataset:}}

We begin with the \textit{Code-to-Text} dataset from the CodeXGLUE benchmark \cite{lu2021codexglue}, from where we first collect pairs $\langle$~\texttt{code, comment}~$\rangle$ including Java and Python. Each instance consists of a code method paired with a human-written natural language summary drawn from large-scale documentation.

Following the original SIDE framework, we reformulated the data for contrastive learning by constructing labeled triplets. The developer-written summary paired with its method serves as the positive sample, yielding 251,820 code–summary pairs. Negative samples were created by randomly pairing each method with a non-matching summary from the training set. This process produces 251,820 ⟨method, positive, negative⟩ triplets, where each method is linked to both an aligned and a non-aligned summary for training the contrastive model.

\subsubsection{\textbf{Model Training with Contrastive Learning:}}

The intuition behind Mastropaolo \etal’s SIDE metric is that contrastive learning maps method-comment pairs onto a latent space where aligned pairs are close and misaligned ones farther apart. This yields an embedding space where distance reflects semantic alignment, enabling the metric to approximate human judgment and complement traditional n-gram–based measures.

Following the SIDE framework, we set up a training objective that minimizes the distance between the embeddings of the code and its corresponding (positive) summary while maximizing the distance between the code and the unrelated (negative) summary. 

We fine-tune MPNet using this objective on our labeled dataset of code-summary pairs. The model encodes both code snippets and their associated summaries independently, and the learned representations are trained to respect the contrastive constraints imposed by the triplets. This process enables the model to capture fine-grained semantic correspondence, serving as the backbone for evaluating summary quality in downstream tasks. 

\subsection{Evaluation}

To evaluate \sidep, we adapted the Python set of the dataset introduced by Crupi \etal \cite{crupi2025effectiveness}, which contains human judgments for a total of 1,163 code summaries across Java and Python. Their dataset was constructed through a rigorous curation process designed to ensure high-quality, human-validated evaluation data.
Specifically, the authors \cite{crupi2025effectiveness} selected the 100 longest functions from each language (Java and Python) in the CoderEval benchmark \cite{yu2024codereval}, based on the number of statements in each function. 
For each function, they collected summaries from two sources: (i) the original developer-written summary, and (ii) summaries automatically generated by five LLMs, including CodeLlama (7B, 13B, 34B), GPT-3.5-turbo, and GPT-4-turbo. 
Each summary was evaluated by three independent human judges, resulting in a total of 3,489 human judgments.
Human evaluation followed a structured rubric inspired by Roy \etal \cite{roy2021reassessing}, assessing summaries along three dimensions: content adequacy, conciseness, and fluency, each rated on a 5-point Likert scale. 
We use this rich, human-annotated dataset as a ground truth to evaluate how well \sidep correlates with human assessments of summary quality, providing a robust basis for validating the metric’s semantic alignment capabilities in the Python domain.
Before delving into the interpretation of our results, we clarify that this section focuses exclusively on the Content Adequacy (CA) dimension, as this is the most directly aligned with the semantic alignment objective of the \sidep metric. Results for Conciseness and Fluency are made available in our replication package\footnote{\url{https://github.com/gitpromptproject/optimizing-data-for-Code-Summarization.git}}.

\begin{table}[ht]
\centering
\caption{Logistic regression results for Content Adequacy (CA). 
Odds Ratios (OR), coefficient values, standard errors, t-values, and p-values are shown for each metric predictor. 
Rows with statistically significant predictors ($p < 0.05$) are shaded. 
The yellow highlight indicates the OR for SIDE score.}
\label{tab:ca}
\small
\resizebox{0.6\textwidth}{!}{%
\begin{tabular}{lccccc}
\toprule
\textbf{Metric} & \textbf{OR} & \textbf{Value} & \textbf{Std. Error} & \textbf{t-value} & \textbf{p-value} \\
\midrule
\rowcolor{gray!20}
BLEU-1 & 0.823 & -0.195 & 0.088 & -2.218 & 0.027 \\
ROUGE-1-P & 0.819 & -0.199 & 0.113 & -1.765 & 0.077 \\
ROUGE-W-R & 1.169 & 0.156 & 0.127 & 1.230 & 0.219 \\
\rowcolor{gray!20}
CodeT5+ CS & 1.526 & 0.422 & 0.182 & 2.320 & 0.020 \\
BERTScore-R & 0.840 & -0.174 & 0.112 & -1.555 & 0.120 \\
\rowcolor{gray!20}
SentenceBERT CS & 1.323 & 0.280 & 0.138 & 2.037 & 0.042 \\
InferSent CS & 1.331 & 0.286 & 0.221 & 1.294 & 0.196 \\
\rowcolor{gray!20}
SIDE score & \cellcolor{yellow!40}\textbf{1.977} & 0.681 & 0.170 & 4.000 & $<0.001$ \\
\rowcolor{gray!20}
C-Coeff & 0.781 & -0.248 & 0.115 & -2.145 & 0.032 \\
\rowcolor{gray!20}
GPT emb & 1.845 & 0.613 & 0.139 & 4.397 & $<0.001$ \\
\bottomrule
\end{tabular}%
}
\end{table}

As in the original SIDE study, we interpret the Odds Ratios (ORs) associated with statistically significant predictors (\ie $p$-value $<$ 0.05), which are shaded in black in \tabref{tab:ca}.

The \sidep score (\texttt{SIDE\_score}) shows a strong and statistically significant effect on CA ratings, with an OR of 1.977 and a $p$-value $<$ 0.001. This means that a one-unit increase in the \sidep score corresponds to an approximate 98\% increase in the odds of observing a higher CA label assigned by human annotators. This is the highest OR among all metrics considered, indicating that \sidep is the strongest predictor of content adequacy in our setting.
The metric also achieves a relatively high regression coefficient (0.6813) and $t$-value (4.0), further supporting its explanatory power. These findings suggest that \sidep effectively captures semantically meaningful aspects of summaries that human evaluators deem content-adequate.
To address RQ\textsubscript{2}, we compare the strength of \sidep correlation with human judgments against other traditional and embedding-based metrics. Among these, \texttt{GPT\_emb} and \texttt{CodeT5\_plus\_CS} also show statistically significant associations ($p$-value $<$ 0.001 and $p$-value = 0.0200, respectively), with ORs of 1.8453 and 1.5798. However, the \sidep metric remains the most effective, not only in terms of statistical significance and OR, but also in task alignment and interpretability--being explicitly designed for evaluating semantic overlap in code summarization tasks. To conclude, \sidep demonstrates the strongest alignment with human-perceived content adequacy, confirming Mastropaolo \etal's results and outperforming both traditional and neural baselines. 

\vspace{5pt}
\begin{tcolorbox}[
    colback=cyan!8, 
    colframe=black, 
    coltext=black,
    arc=6pt, 
    boxrule=0.8pt, 
    left=5pt, 
    right=5pt, 
    top=8pt, 
    bottom=8pt, 
    fonttitle=\bfseries,
    coltitle=black, 
    title=Study 2 Summary,
    enhanced,
    attach boxed title to top left={yshift=-3mm, xshift=5mm},
    boxed title style={
        colback=gray!40,
        boxrule=0.7pt,
        arc=8pt,
        outer arc=8pt,
        left=5pt,
        right=5pt,
        top=0.5pt,
        bottom=0.5pt,
    }
]
\sidep has been proven to be a valid semantic alignment metric for Python code summarization. Specifically,
\sidep achieves the strongest correlation with human-perceived content adequacy among all evaluated metrics. 
Compared to alternative approaches, \sidep outperforms both traditional lexical metrics (BLEU, ROUGE) and embedding-based alternatives (GPT embeddings: OR = 1.8453; CodeT5+ cosine similarity: OR = 1.5798). 
Our findings confirm that contrastive-based semantic alignment metrics must be calibrated to language-specific syntactic and naming conventions rather than assumed transferable across programming languages. 
\end{tcolorbox}


\section{Study 3: Differentiated Replication of Study 1 on Python}
\label{sec:study-3}

Finally, we investigate the effectiveness of previously introduced dataset optimization strategies---semantic alignment-based and token-based---in the context of Python code summarization. The primary objective is to evaluate the generalizability of techniques originally developed for Java datasets and determine whether they remain effective when applied to a different programming language.
To apply semantic alignment optimization, we adopt the filtering pipeline proposed by Vitale \etal, incorporating our retrained Python-specific metric, \sidep. Consistent with the experimental design outlined in Study 1 (\secref{sec:study-1}), we evaluate the optimization strategies on both--independently and in combination with \sidep--on the Python dataset.
To enable direct comparison and continuity with prior experiments, we address a parallel set of RQs as those defined in Study 1. For brevity, we do not restate them here.

\subsection{Dataset Preparation and Optimization}
To initiate this investigation, we first replicated the semantic alignment-based filtering pipeline from Vitale \etal using \sidep. As the training corpus, we selected the Python version of the Funcom dataset, originally released by Su \etal \cite{su2024semantic}. In their study, Vitale \etal utilized the cleaned version of the Funcom Java dataset, curated by Shi \etal~\cite{shi2022we} using their proposed tool CAT (Code-Comment Cleaning Tool). Since only the Java version of Funcom was cleaned in the original study, to align with Study 1, we applied CAT to the unoptimized Python Funcom dataset (250,000 instances) to produce a consistent and high-quality training set. This cleaning step resulted in a curated dataset of 247,695 $\langle$code, comment$\rangle$ pairs. Following Vitale \etal's approach, we then applied \sidep-based semantic filtering with a threshold of 0.9, yielding a final semantically-aligned dataset of 78,611 instances, which served as the foundation for this study.


To address RQ1, we applied the three token-based optimization techniques introduced in Study 1, following the same implementation pipeline. \cbleuApproach achieves 72.23\% token reduction, \functionApproach extraction yields 71.35\% reduction, and \astApproach-based transformation produces 55.69\% token reduction on the unoptimized Fncom\textsubscript{Python} dataset.

For RQ2, we replicated the semantic filtering strategy using the \sidep metric, applying a threshold of 0.9--consistent with the configuration in Study 1--to filter low-alignment instances. This resulted in a semantically refined \sidep optimized training dataset. Next, we applied token-based optimization on top of the \sidep-filtered data to evaluate whether combining both strategies yields further improvements. All token-level reduction techniques were implemented as described in \secref{subsec:data-op} of Study 1.

These processes yielded eight distinct training datasets, each representing a different combination of optimization strategies. We fine-tuned eight separate models---one per dataset---on the code summarization task using the same training configuration outlined in Study 1 \secref{subsec:model-training}.

\subsection{Experiment and Evaluation}

\subsubsection{\textbf{Evaluation Dataset:}}
To assess the effectiveness of the optimized models, we curated a high-quality Python benchmark for evaluating code summarization. This benchmark consists of 500 human-annotated instances selected to represent high-quality $\langle$code, summary$\rangle$ pairs.
Additionally, we utilize the CoderEval Python dataset \cite{yu2024codereval} as a secondary evaluation benchmark. Following the same adaptation approach described in Study 1 for Java, we inverted the original $\langle$ \texttt{docstring, code} $\rangle$ format to $\langle$ \texttt{code, docstring} $\rangle$ to enable code summarization evaluation. To ensure data quality, we performed manual inspection following the same procedure described by Vitale \etal: two authors independently reviewed all instances to verify code-docstring alignment. We extracted only the first sentence from multi-sentence docstrings to maintain consistency with the single-sentence summary format and excluded those containing mathematical formulas not explained in natural language. This curation process resulted in a final evaluation set of 230 high-quality Python code-summary pairs, providing a complementary assessment of our token optimization strategies across different evaluation contexts.

\subsubsection{\textbf{Benchmark data construction:}}
To ensure the integrity of the benchmark and minimize potential data contamination from pretraining corpora, we selected only GitHub repositories created after August 2024---the release date of GPT-4o-mini \cite{gpt4} (our evaluation model). Additional filters were applied to ensure repository quality, including a minimum of 100 stars, 100 commits, and at least one fork. From these repositories, we extracted 25,000 Python code-comment pairs and applied the CAT tool \cite{shi2022we} to remove noisy examples, yielding 8,675 cleaned $\langle comment, code \rangle$ pairs.
Subsequently, we filtered the data to retain only those instances where the code contained 20 to 200 tokens \cite{liu2019neural, bifolco2025llms} and the corresponding summary was longer than three tokens. The final benchmark consists of 500 curated examples. To generate system summaries for evaluation, we used GPT-4o-mini \cite{gpt4}, a state-of-the-art LLM.


To create gold-standard annotations, two authors manually rated the generated summaries based on three quality dimensions--Content Adequacy, Fluency, and Conciseness--using the evaluation rubric proposed by Roy \etal~\cite{roy2021reassessing}. Disagreements were resolved through discussion with a third author to ensure inter-annotator reliability. This rigorous annotation process yielded a high-quality evaluation benchmark comprising 500 manually curated Python code-summary pairs, each accompanied by human quality assessments across the three dimensions. These human-validated annotations serve as the gold-standard ground truth for evaluating the performance of models trained using different optimization strategies investigated in Study 3, enabling a reliable assessment of how token-level optimization techniques affect summarization quality in real-world scenarios. We refer to this new benchmark as \textit{PyBench} throughout the remainder of this paper.

\subsubsection{\textbf{Evaluation Metrics}}
To assess the effectiveness of our optimization strategies on Python datasets, we employed the same comprehensive evaluation framework established in Study 1 (\secref{sec:study-1}), which included BLEU, ROUGE, and METEOR metrics for summarization quality assessment, alongside token retention ratio and Shannon entropy reduction to measure optimization effectiveness.



\subsubsection{\textbf{Results:}}

\begin{table*}[t]
\centering
\caption{Evaluation of \textbf{standalone token-optimization} strategies on the Funcom\textsubscript{Python} dataset using CodeT5+ (220M). The second column shows the actual entropy value after optimization, with parenthesized values indicating the percentage of entropy reduction. Best scores per metric are in \textbf{bold}.}
\label{tab:res-3_2}
\footnotesize
\resizebox{\textwidth}{!}{%
\begin{tabular}{lccccccccc}
\toprule
\multirow{3}{*}{\textbf{Approach}} & 
\multirow{3}{*}{\textbf{\begin{tabular}[c]{@{}c@{}}Entropy Reduction\\ (vs. Original)\end{tabular}}} & 
\multirow{3}{*}{\textbf{\begin{tabular}[c]{@{}c@{}}Token\\Retention\end{tabular}}} &
\multicolumn{3}{c}{\textbf{CoderEval-Python\cite{su2024semantic}}} & 
\multicolumn{3}{c}{\textbf{\textit{PyBench}}} \\
\cmidrule(lr){4-6} \cmidrule(lr){7-9}
& & & \textbf{BLEU} & \textbf{ROUGE-L} & \textbf{METEOR} & \textbf{BLEU} & \textbf{ROUGE-L} & \textbf{METEOR} \\
\midrule
Original Funcom\textsubscript{Python} & 4.448 & 100\% & \textbf{2.60} & \textbf{21.44} & \textbf{12.56} & 1.74 & 16.32 & 8.68 \\
\midrule
\cbleuApproach & 3.482 (\textcolor{red}{$\downarrow$}21.72\%) & 39.24\% & 2.08 & 16.69 & 9.56 & 1.95 & 15.63 & 9.01 \\
\midrule
\functionApproach & \bf \cellcolor{yellow!60!orange!30}3.241 (\textcolor{red}{$\downarrow$}27.14\%) & 16.73\% & 2.45 & 21.33 & 12.37 & \textbf{2.43} & \textbf{20.19} & \textbf{11.08} \\
\midrule
\astApproach & 3.852 (\textcolor{red}{$\downarrow$}13.39\%) & 43.50\% & 1.69 & 13.26 & 7.39 & 0.58 & 4.35 & 2.50 \\
\midrule
\rowcolor{blue!10} 
SIDE Optimized Funcom\textsubscript{Python} & 
\royalval{4.453} & 
\royalval{--} &
\royalval{3.07} & 
\royalval{23.09} & 
\royalval{13.97} & 
\royalval{2.69} & 
\royalval{21.67} & 
\royalval{12.45} \\
\bottomrule
\end{tabular}%
}
\end{table*}

\begin{table*}[t]
\centering
\caption{Evaluation of \textbf{token-optimization} strategies on the \textbf{SIDE Optimized} Funcom\textsubscript{Python} dataset using CodeT5+ (220M). The second column shows the entropy for each optimization technique, with parenthesized values indicating the percentage of entropy reduction. Best scores per metric are in \textbf{bold}.}
\label{tab:res-3_1}
\footnotesize
\resizebox{\textwidth}{!}{%
\begin{tabular}{lccccccccc}
\toprule
\multirow{3}{*}{\textbf{Approach}} & 
\multirow{3}{*}{\textbf{\begin{tabular}[c]{@{}c@{}}Entropy Reduction\\ (vs. Original)\end{tabular}}} & 
\multirow{3}{*}{\textbf{\begin{tabular}[c]{@{}c@{}}Token\\Retention\end{tabular}}} &
\multicolumn{3}{c}{\textbf{CoderEval-Python\cite{su2024semantic}}} & 
\multicolumn{3}{c}{\textbf{\textit{PyBench}}} \\
\cmidrule(lr){4-6} \cmidrule(lr){7-9}
& & & \textbf{BLEU} & \textbf{ROUGE-L} & \textbf{METEOR} & \textbf{BLEU} & \textbf{ROUGE-L} & \textbf{METEOR} \\
\midrule
SIDE Optimized Funcom\textsubscript{Python} & 4.453 & 100\% & \textbf{3.07} & \textbf{23.09} & \textbf{13.97} & 2.69 & \textbf{21.67} & \textbf{12.45} \\
\midrule
\cbleuApproach & 3.475 (\textcolor{red}{$\downarrow$}21.97\%) & 37.14\% & 2.65 & 18.88 & 11.87 & 1.99 & 14.84 & 8.43 \\
\midrule
\functionApproach & \bf \cellcolor{yellow!60!orange!30}3.264 (\textcolor{red}{$\downarrow$}26.71\%) & 16.93\% & 2.81 & 22.30 & 13.96 & \textbf{2.78} & 20.96 & 12.21 \\
\midrule
\astApproach & 3.869 (\textcolor{red}{$\downarrow$}13.11\%) & 43.60\% & 1.56 & 11.09 & 5.83 & 0.59 & 4.35 & 2.35 \\
\bottomrule
\end{tabular}%
}
\end{table*}

The results in \tabref{tab:res-3_2} and \tabref{tab:res-3_1} reveal striking patterns in how token-level optimization strategies perform across programming languages, with Shannon entropy and token retention serving as our measures of compression effectiveness.

Examining the \textbf{standalone token-optimization} results in \tabref{tab:res-3_1}, we observe a markedly different performance landscape compared to Java, where, we remind the reader that average BLEU, ROUGE-L, and METEOR scores are consistently higher.
Function Signature extraction achieves the most aggressive compression with 27.14\% entropy reduction and only 16.73\% \reductionicon{16.73} token retention. Remarkably, this extreme optimization achieves the best performance among the three approaches, yielding a BLEU score of 2.45 on CoderEval-Python (94\% of the 2.60 baseline) and 2.43 on our newly proposed benchmark---\textit{PyBench} (140\% of the baseline), alongside a ROUGE-L score of 21.33 and a METEOR score of 12.37. The performance differences are minimal, with \functionApproach showing only marginal losses of -0.011 in ROUGE-L and -0.019 in METEOR.

On \textit{PyBench}, providing the model with only Python signatures as input also yields optimal results. We attribute this to the higher quality level of our benchmark resulting from the filtering process. This finding inverts the pattern observed in Java: while Function Signatures showed the steepest quality trade-offs for Java, they prove most effective for Python. This contrast suggests that Python's more standardized and semantically rich function signatures contain sufficient information for accurate summarization, whereas Java's more verbose syntax may require additional structural context.

CrystalBLEU provides moderate compression with 21.72\% entropy reduction, retaining 39.24\% \reductionicon{39.24} of tokens. On CoderEval-Python, this yields a BLEU of 2.08 (80\% of baseline), while on \textit{PyBench} it achieves 1.95--a notable improvement over the challenging 1.74 baseline (112\%). The consistent pattern of improvement on the harder \textit{PyBench} dataset reinforces the observation from Java: aggressive token reduction can actually enhance model performance when dealing with complex or poorly documented code by helping models focus on essential signals rather than noise.

By contrast, AST-based optimization shows significantly different behavior in Python. Despite the most conservative compression (13.39\% entropy reduction, 43.50\% \reductionicon{43.50} token retention), AST achieves the weakest performance--BLEU of 1.69 on CoderEval-Python and a striking 0.58 on \textit{PyBench} (only 33\% of baseline). The large degradation on \textit{PyBench} suggests that Python's AST representation may strip away critical contextual information that the model needs for effective summarization. This contrasts sharply with Java, where AST consistently delivered the strongest results.
Statistical validation of these patterns is presented in \tabref{tab:statistical-st3}, with the top part showing SIDE-optimized comparisons and the bottom part showing standalone comparisons.

The \textbf{cascaded results} in \tabref{tab:res-3_2} demonstrate how semantic pre-filtering interacts with token optimization in Python. Starting from a stronger SIDE-optimized baseline (BLEU 3.07 on CoderEval-Python, 2.69 on \textit{PyBench}), we observe that Function Signature extraction maintains its effectiveness--achieving 2.81 BLEU on CoderEval (92\% of baseline) with 26.71\% entropy reduction and only 16.93\% \reductionicon{16.93} token retention. On \textit{PyBench}, it delivers 2.78 BLEU (103\% of baseline) while substantially improving ROUGE-L (20.96 vs. 21.67, maintaining 97\%) and METEOR (12.21 vs. 12.45, preserving 98\%). 

CrystalBLEU shows similar patterns with 21.97\% entropy reduction (retaining 37.14\% \reductionicon{37.14} of tokens), achieving competitive performance on CoderEval (BLEU: 2.65, 86\% of baseline) and maintaining quality on \textit{PyBench} (BLEU: 1.99, 74\% of baseline). However, AST continues to struggle--BLEU drops to 1.56 on CoderEval (51\% of baseline) and collapses to 0.59 on \textit{PyBench}, with ROUGE-L falling to 4.35 from 21.67 (80\% degradation). This consistent AST underperformance across both standalone and cascaded settings, on both evaluation benchmarks, reveals a fundamental incompatibility between Python's AST representation and the summarization task as implemented in our framework.

A key insight emerges when comparing the effectiveness of token optimization techniques across equivalent tasks in the CoderEval benchmark--where the same problems are implemented once in Python and once in Java (\tabref{tab:cross-lang-comparison}). The findings reveal that the success of these strategies is strongly language-dependent. In Java, AST-based optimization achieved up to a 37\% improvement over the baseline, whereas in Python it performed poorly, even degrading output quality. Conversely, Function Signatures incurred notable quality trade-offs in Java but delivered the strongest results in Python, improving performance even on more challenging datasets. These observations suggest that the semantic density and structural characteristics of code representations differ fundamentally between programming languages--likely driven by variations in syntax verbosity, naming conventions, and documentation practices.


\begin{table*}[t]
\centering
\caption{Cross-language comparison of token optimization effectiveness on CoderEval benchmarks. For each metric, the results for Java and Python are presented side by side. Best scores per language among optimization methods are in bold.}
\label{tab:cross-lang-comparison}
\tiny 
\resizebox{0.8\textwidth}{!}{%
\begin{tabular}{l>{\columncolor{white}}c cccccccc}
\toprule
& & \multicolumn{2}{c}{\textbf{BLEU}} & \multicolumn{2}{c}{\textbf{ROUGE-L}} & \multicolumn{2}{c}{\textbf{METEOR}} & \multicolumn{2}{c}{\textbf{Token Retention}} \\
\cmidrule(lr){3-4} \cmidrule(lr){5-6} \cmidrule(lr){7-8} \cmidrule(lr){9-10}
\textbf{Method} & \textbf{Optimization} & Java & Python & Java & Python & Java & Python & Java & Python \\
\midrule
\multirow{2}{*}{Baseline} 
& Original & 12.41 & 2.60 & 36.94 & 21.44 & 35.35 & 12.56 & 100\% & 100\% \\
& \cellcolor{gray!30}SIDE-Opt & \cellcolor{gray!30}12.13 & \cellcolor{gray!30}3.07 & \cellcolor{gray!30}42.15 & \cellcolor{gray!30}23.09 & \cellcolor{gray!30}35.37 & \cellcolor{gray!30}13.97 & \cellcolor{gray!30}100\% & \cellcolor{gray!30}100\% \\
\midrule
\multirow{2}{*}{CrystalBLEU} 
& Standalone & 10.48 & 2.08 & 31.09 & 16.69 & 40.90 & 9.56 & 27.65\% & 39.24\% \\
& \cellcolor{gray!30}SIDE-Opt & \cellcolor{gray!30}9.56 & \cellcolor{gray!30}2.65 & \cellcolor{gray!30}28.07 & \cellcolor{gray!30}18.88 & \cellcolor{gray!30}37.89 & \cellcolor{gray!30}11.87 & \cellcolor{gray!30}27.77\% & \cellcolor{gray!30}37.14\% \\
\midrule
\multirow{2}{*}{Function Signature} 
& Standalone & 8.10 & \textbf{2.45} & 21.86 & \textbf{21.33} & 33.78 & \textbf{12.37} & 30.49\% & 16.73\% \\
& \cellcolor{gray!30}SIDE-Opt & \cellcolor{gray!30}10.81 & \cellcolor{gray!30}\textbf{2.81} & \cellcolor{gray!30}32.94 & \cellcolor{gray!30}\textbf{22.30} & \cellcolor{gray!30}41.72 & \cellcolor{gray!30}\textbf{13.96} & \cellcolor{gray!30}28.65\% & \cellcolor{gray!30}16.93\% \\
\midrule
\multirow{2}{*}{AST} 
& Standalone & 9.16 & 1.69 & 26.49 & 13.26 & 37.62 & 7.39 & 44.08\% & 43.50\% \\
& \cellcolor{gray!30}SIDE-Opt & \cellcolor{gray!30}\textbf{11.81} & \cellcolor{gray!30}1.56 & \cellcolor{gray!30}\textbf{38.01} & \cellcolor{gray!30}11.09 & \cellcolor{gray!30}\textbf{45.45} & \cellcolor{gray!30}5.83 & \cellcolor{gray!30}44.31\% & \cellcolor{gray!30}43.60\% \\
\bottomrule
\end{tabular}%
}
\end{table*}

\begin{table*}[t]
\centering
\caption{Statistical comparison of approaches for Python SIDE-Optimized and Standalone Dataset. P-values are reported as ``x'' when $\geq$ 0.05 (not statistically significant) and as the actual value when $< 0.05$ (significant). Cliff's Delta effect sizes are provided for all comparisons with labels indicating magnitude: N (Negligible: $|\delta| < 0.147$), S (Small: $0.147 \leq |\delta| < 0.33$), M (Medium: $0.33 \leq |\delta| < 0.474$), L (Large: $|\delta| \geq 0.474$).}
\label{tab:statistical-st3}
\footnotesize
\resizebox{\textwidth}{!}{%
\begin{tabular}{llcccccccccccc}
\toprule
\multirow{3}{*}{\textbf{Dataset}} & \multirow{3}{*}{\textbf{Metrics}} & \multicolumn{6}{c}{\textbf{CoderEval-Python \cite{yu2024codereval}}} & \multicolumn{6}{c}{\textbf{\textit{PyBench}}} \\
\cmidrule(lr){3-8} \cmidrule(lr){9-14}
& & \multicolumn{2}{c}{\textbf{CrystalBLEU}} & \multicolumn{2}{c}{\textbf{Function Signature}} & \multicolumn{2}{c}{\textbf{AST}} & \multicolumn{2}{c}{\textbf{CrystalBLEU}} & \multicolumn{2}{c}{\textbf{Function Signature}} & \multicolumn{2}{c}{\textbf{AST}} \\
\cmidrule(lr){3-4} \cmidrule(lr){5-6} \cmidrule(lr){7-8} \cmidrule(lr){9-10} \cmidrule(lr){11-12} \cmidrule(lr){13-14}
& & \textbf{p-value} & \textbf{Cliff's $\delta$} & \textbf{p-value} & \textbf{Cliff's $\delta$} & \textbf{p-value} & \textbf{Cliff's $\delta$} & \textbf{p-value} & \textbf{Cliff's $\delta$} & \textbf{p-value} & \textbf{Cliff's $\delta$} & \textbf{p-value} & \textbf{Cliff's $\delta$} \\
\midrule
\multirow{3}{*}{SIDE-Optimized Funcom\textsubscript{Python}} 
& BLEU & x & -0.053 (N) & x & 0.015 (N) & <0.001 & -0.420 (M) & <0.001 & -0.215 (S) & x & 0.001 (N) & <0.001 & -0.611 (L) \\
& ROUGE-L & <0.001 & -0.196 (S) & x & -0.014 (N) & <0.001 & -0.514 (L) & <0.001 & -0.303 (S) & x & -0.035 (N) & <0.001 & -0.729 (L) \\
& METEOR & 0.003 & -0.127 (N) & x & 0.027 (N) & <0.001 & -0.518 (L) & <0.001 & -0.287 (S) & x & -0.027 (N) & <0.001 & -0.715 (L) \\
\midrule
\multirow{3}{*}{Standalone Token-Optimized Funcom\textsubscript{Python}} 
& BLEU & 0.010 & -0.110 (N) & x & -0.025 (N) & <0.001 & -0.235 (S) & x & 0.024 (N) & <0.001 & 0.135 (N) & <0.001 & -0.427 (M) \\
& ROUGE-L & <0.001 & -0.205 (S) & x & 0.003 (N) & <0.001 & -0.366 (M) & x & -0.053 (N) & <0.001 & 0.160 (S) & <0.001 & -0.576 (L) \\
& METEOR & <0.001 & -0.169 (S) & x & 0.006 (N) & <0.001 & -0.314 (S) & x & -0.017 (N) & <0.001 & 0.152 (S) & <0.001 & -0.530 (L) \\
\bottomrule
\end{tabular}%
}
\end{table*}
To validate these observations, we conducted \textbf{Wilcoxon signed-rank} tests with Cliff's delta effect size measurements (\tabref{tab:statistical-st3}), comparing each optimization method against its respective baseline using paired tests. For SIDE-optimized data, we compared each token method against the SIDE-filtered baseline. For standalone optimizations, we compared the results against the original, unoptimized dataset. To account for multiple comparisons across three methods and three metrics, we applied Holm-Bonferroni correction to adjust p-values and control the familywise error rate. The statistical patterns reveal striking differences from those in Java.

On SIDE-optimized data, AST shows large negative effects across all metrics on both benchmarks (Cliff's delta: -0.420 to -0.729 on \textit{PyBench}, -0.420 to -0.518 on CoderEval-Python), with all comparisons highly significant (p $<$ 0.05), confirming that AST degradation is statistically robust and practically severe. Function Signatures, by contrast, show negligible effects on CoderEval-Python (Cliff's delta: -0.014 to 0.027, all p $>$ 0.05) and negligible effects on \textit{PyBench} (Cliff's delta: -0.035 to 0.001, all p $>$ 0.05), indicating that aggressive compression through signatures maintains statistical equivalence with baselines despite 83\% token removal. CrystalBLEU demonstrates small-to-medium negative effects on both benchmarks (Cliff's delta: -0.127 to -0.303), with statistical significance on most comparisons.

For standalone optimization, the patterns intensify. AST continues to show medium-to-large negative effects (Cliff's delta: -0.235 to -0.576), with all comparisons highly significant (p $<$ 0.05), reinforcing its unsuitability for Python summarization. Function Signatures demonstrate a remarkable reversal: on \textit{PyBench}, they show small positive effects (Cliff's delta: 0.135 to 0.160) with statistical significance across all metrics (p $<$ 0.001), confirming that this method actually improves summary quality on challenging datasets. On CoderEval-Python, Function Signatures maintain negligible effects (Cliff's delta: -0.025 to 0.006). CrystalBLEU shows small negative effects across metrics (Cliff's delta: -0.110 to -0.205), with most comparisons achieving significance.

These empirical findings yield critical insights for cross-language code summarization. First, token optimization strategies are not language-agnostic--a method that excels in one language may catastrophically fail in another. The AST-Python mismatch, confirmed by large negative effect sizes across all experimental conditions, demonstrates that structural representations must be carefully validated for each language rather than assumed to be transferable. Second, Python's concise syntax and expressive function signatures enable more aggressive compression than Java. Function Signatures achieve 83\% token reduction with statistical equivalence or improvement, whereas Java requires more conservative approaches for optimal performance. Third, the consistent pattern where Function Signatures improve performance on the challenging \textit{PyBench} (positive effect sizes, p $<$ 0.001) while maintaining equivalence on cleaner CoderEval data suggests that compression strategies should be matched to corpus characteristics rather than applied uniformly.

Taken together, these patterns suggest clear guidance for Python code summarization systems: Function Signature extraction emerges as the dominant strategy, offering 83\% token reduction while maintaining or improving quality across diverse evaluation contexts. CrystalBLEU provides a fallback option with 60-63\% compression and acceptable quality trade-offs when function signatures alone prove insufficient. AST-based optimization should be avoided for Python summarization, given its consistent catastrophic failures across all experimental conditions. Importantly, these recommendations stand in stark contrast to Java guidance (where AST dominated), underscoring the critical importance of language-specific validation before deploying token optimization strategies in production systems. Complete statistical results are available in our replication package~\cite{replication}.

\vspace{10pt}

\begin{tcolorbox}[
    colback=cyan!8, 
    colframe=black, 
    coltext=black,
    arc=6pt, 
    boxrule=0.8pt, 
    left=5pt, 
    right=5pt, 
    top=8pt, 
    bottom=8pt, 
    fonttitle=\bfseries,
    coltitle=black, 
    title=Study 3 Summary,
    enhanced,
    attach boxed title to top left={yshift=-3mm, xshift=5mm},
    boxed title style={
        colback=gray!40,
        boxrule=0.7pt,
        arc=8pt,
        outer arc=8pt,
        left=5pt,
        right=5pt,
        top=0.5pt,
        bottom=0.5pt,
    }
]

Our findings reveal that token optimization strategies are language-dependent. AST achieves 37\% improvements in Java but suffers 35-49\% performance loss in Python, while Function Signatures show the opposite pattern--poor Java performance but optimal Python results with 83\% token reduction and only 6-8\% degradation. Python's concise syntax enables more aggressive compression 
than Java's verbose structure, which requires conservative approaches. We also found that the compression effectiveness varies with dataset complexity.
In general, the choice of the optimal strategy 
must be calibrated to language-specific characteristics (\tabref{tab:cross-lang-comparison}). 
\end{tcolorbox}

\section{Qualitative Examples} \label{sec:qa-implications}


To illustrate the practical impact of token optimization strategies beyond aggregate metrics, we present a representative example of code summarization from our Python evaluation that demonstrates how different token reduction methods fundamentally alter the semantic content captured by trained models.

\begin{figure}[htbp]
\centering
\includegraphics[width=0.5\columnwidth]{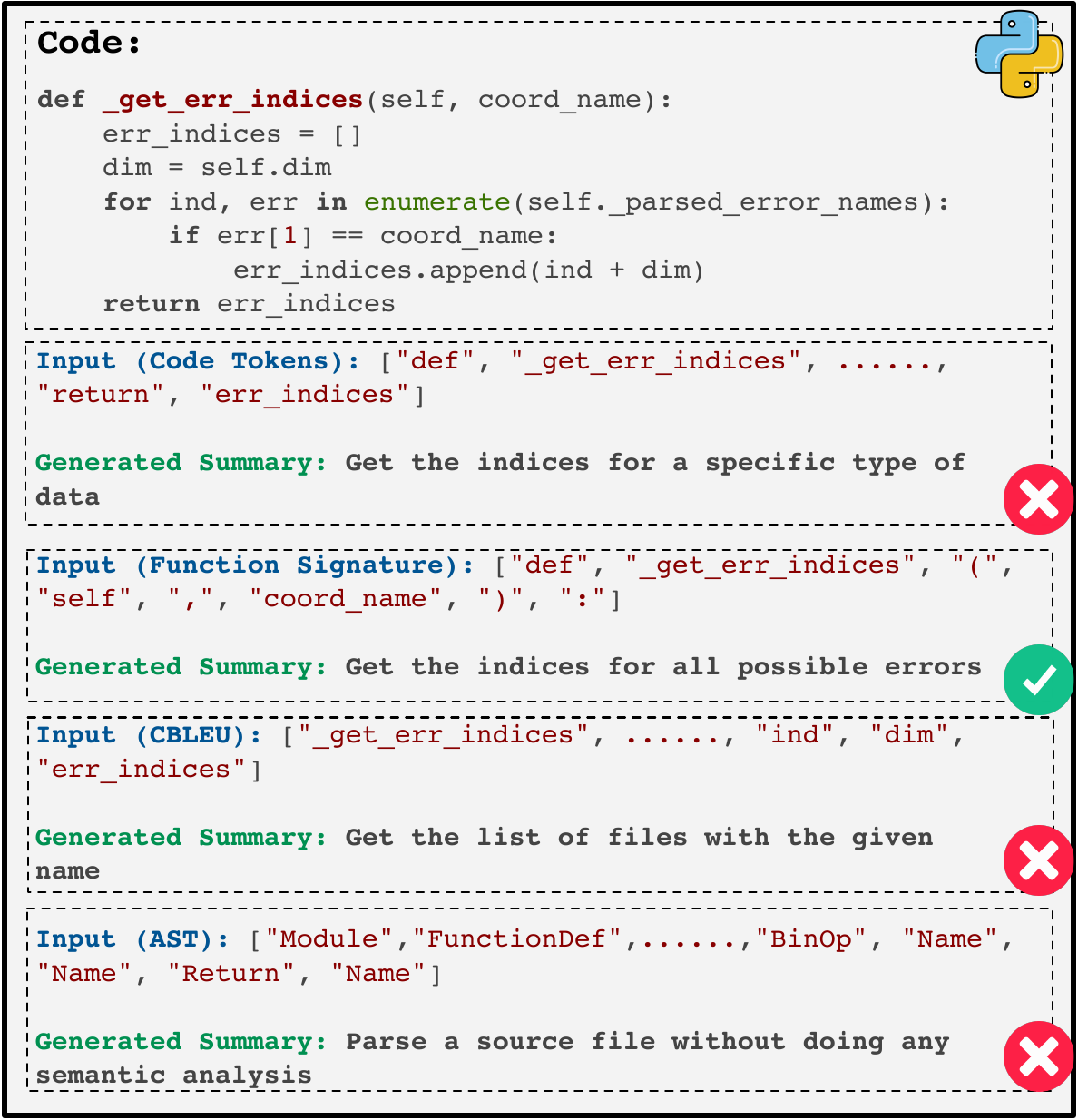}
\caption{Qualitative comparison of token optimization strategies on Python code summarization. Function Signature extraction (\textcolor{green}{\ding{51}}) successfully captures the core semantics, generating an accurate summary, while full code tokens, CrystalBLEU, and AST representation (\textcolor{red}{\ding{55}}) all produce irrelevant summaries that misidentify the function's purpose.}
\label{fig:quali-py}
\end{figure}

\figref{fig:quali-py} shows the function \texttt{get\_err\_indices(self, coord\_name)}, which iterates through parsed error names, checks if the error matches a given coordinate name, and returns the indices where matches occur. This function exemplifies common Python patterns: descriptive naming, iteration with conditionals, and list manipulation.

\setlength{\textfloatsep}{10pt}

When trained on full code tokens, the model generates: ``Get the indices for a specific type of data''--a vague, generic summary that captures the high-level concept of retrieving indices but misses the critical detail that the function identifies error indices based on coordinate name matching. This demonstrates a common failure: despite having access to complete implementation details (\texttt{enumerate}, \texttt{self.\_parsed\_error\_names}, conditional matching \texttt{err[1] == coord\_name}), the model produces an overly abstract summary that loses functional specificity. The verbosity of full code tokens appears to dilute the semantic signal, causing the model to retreat to safe, generic patterns.

Function Signature extraction, retaining only \texttt{["def", "\_get\_err\_indices", "(", "self", ",", "coord\_name", ")", ":"]} (representing the method name and parameters), produces: ``Get the indices for all possible errors''--the most semantically accurate summary despite extreme token reduction. The function name \texttt{get\_err\_indices} explicitly encodes the retrieval action and target (error indices), while the parameter \texttt{coord\_name} signals filtering based on coordinate names. This combination proves sufficient for the model to accurately infer the function's purpose. The slight generalization to ``all possible errors'' demonstrates reasonable semantic inference from limited but highly informative tokens. This example illustrates why Function Signatures excel in Python: descriptive naming conventions and meaningful parameter names encode functional intent that proves more valuable than implementation details.

CrystalBLEU optimization, retaining tokens like \texttt{["\_get\_err\_indices", ......, "ind", "dim", "err\_indices"]}, generates: ``Get the list of files with the given name''--a semantically plausible but functionally incorrect summary. By preserving variable names (\texttt{ind}, \texttt{dim}, \texttt{err\_indices}) while removing structural keywords (\texttt{def}, \texttt{return}, \texttt{for}, \texttt{if}), CrystalBLEU retains lexical elements but strips away the control flow that defines the function's logic. Without keywords like \texttt{return} indicating what the function produces, or \texttt{for}/\texttt{if} revealing the filtering operation, the model misinterprets the retained tokens as file-related operations--possibly influenced by common patterns where \texttt{get} combined with list-like variables suggests file retrieval. This failure demonstrates that frequency-based pruning can eliminate structurally essential tokens (such as \texttt{return}, \texttt{for}, and \texttt{if}) that, despite being common, provide critical context for understanding program semantics.

AST-based representation, transforming the code into abstract tokens like \texttt{["Module", "FunctionDef", ......, "BinOp", "Name", "Name", "Return", "Name"]}, generates: ``Parse a source file without doing any semantic analysis''--a summary completely unrelated to the actual functionality. The AST preserves high-level syntactic structure (module, function definition, binary operation, return statement) but abstracts away all identifier information. Without access to tokens like \texttt{get\_err\_indices}, \texttt{coord\_name}, or \texttt{err\_indices}, the model cannot determine what the function operates on or returns. The hallucinated functionality (``parse a source file'') likely reflects the model's attempt to match structural patterns (functions with iteration and conditionals) to memorized templates from training data, resulting in a semantically disconnected prediction.


To further illustrate the language-dependent nature of token optimization, we present a representative Java example that demonstrates why AST-based optimization excels in this language while methods that succeeded in Python fail.

\begin{figure}[htbp]
\centering
\includegraphics[width=0.5\columnwidth]{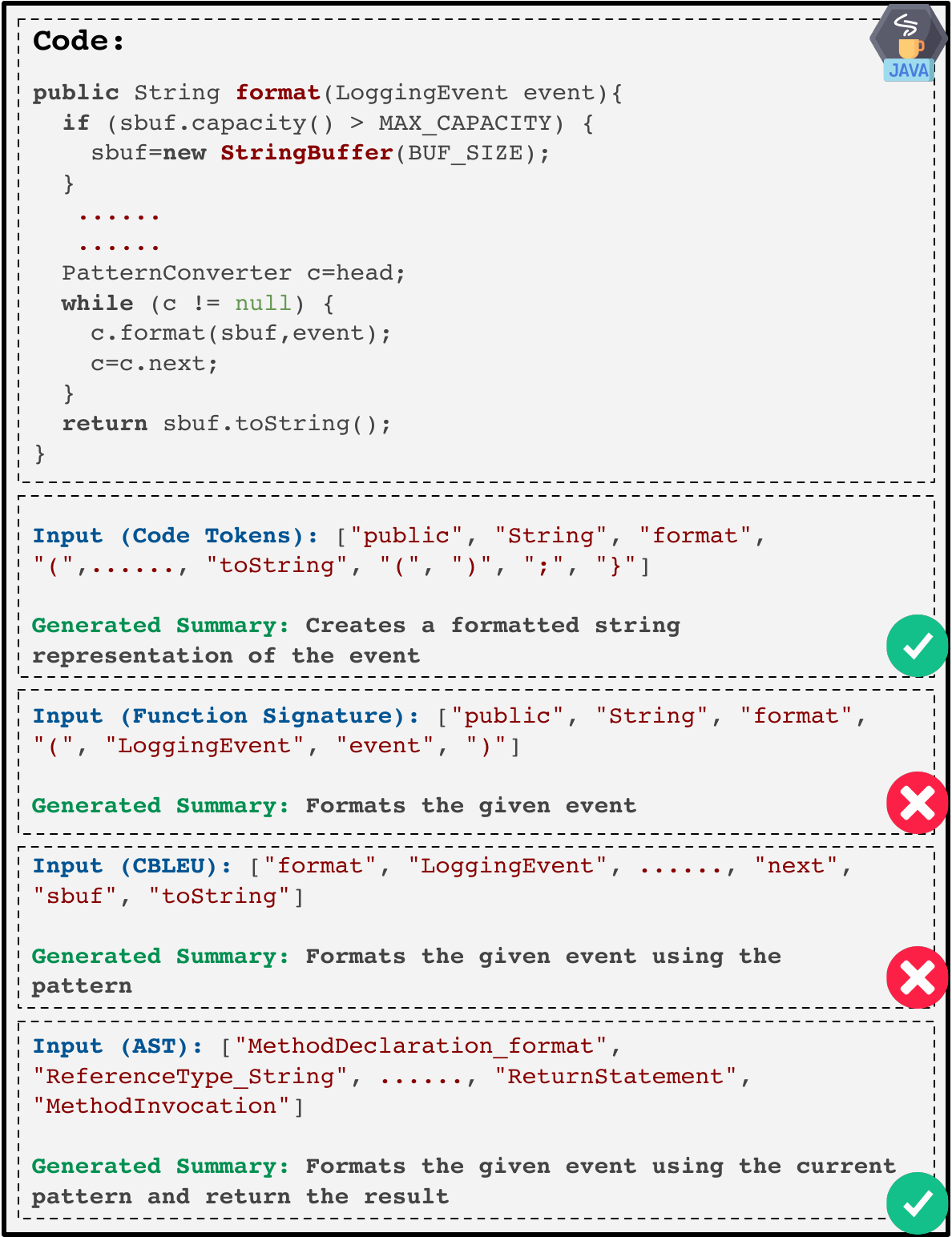}
\caption{Qualitative comparison of token optimization strategies on Java code summarization. Full code tokens (\textcolor{green}{\ding{51}}) and AST representation (\textcolor{green}{\ding{51}}) successfully capture the core functionality, while Function Signature and CrystalBLEU (\textcolor{red}{\ding{55}}) produce incorrect or incomplete summaries.}
\label{fig:quali-java}
\end{figure}

\figref{fig:quali-java} shows the Java method \texttt{format(LoggingEvent event)}, which formats a logging event into a string representation. The method checks buffer capacity, reinitializes if needed, uses a pattern converter chain to format the event, and returns the resulting string. This exemplifies typical Java patterns: explicit type declarations, conditional buffer management, iterator-style traversal, and verbose method calls.

\setlength{\textfloatsep}{10pt}

When trained on full code tokens, the model generates: ``Creates a formatted string representation of the event''---an accurate summary that captures both the formatting action and the target (event). With access to complete implementation details, including type information (\texttt{String}, \texttt{LoggingEvent}), method names (\texttt{format}, \texttt{toString}), and control structures (\texttt{if}, \texttt{while}), the model successfully infers the method's purpose. 

Function Signature extraction, retaining \texttt{["public", "String", "format", "(", "LoggingEvent", "event", ")"]}, generates: ``Formats the given event''--a technically correct but incomplete summary. While the method name \texttt{format} and parameter type \texttt{LoggingEvent} suggest formatting functionality, the signature alone cannot convey that the method creates a string representation or handles buffer management. This demonstrates a key difference from Python: Java method names tend to be less descriptive (generic verbs like \texttt{format}, \texttt{process}, \texttt{handle}), placing less semantic weight on identifiers alone. The explicit return type \texttt{String} helps, but without implementation context, the model produces an overly simplified summary that misses important functional details.

CrystalBLEU optimization, retaining tokens like \texttt{["format", "LoggingEvent", ......, "next", "sbuf", "toString"]}, generates: ``Formats the given event using the pattern''---a summary that adds spurious details. By preserving method names and variable names while removing type declarations and control structures, CrystalBLEU creates ambiguity. The retained token \texttt{pattern} (from \texttt{PatternConverter}) misleads the model into believing pattern-based formatting is the primary feature, when in reality it's an implementation detail. Without a structural context that shows the buffer management and iteration logic, the model focuses on high-frequency domain terms, resulting in partially incorrect inferences about the method's core functionality.

AST-based representation, transforming the code into \texttt{["MethodDeclaration\_format", "ReferenceType\_String", ......, "ReturnStatement", "MethodInvocation"]}, generates: ``Formats the given event using the current pattern and returns the result''--the most complete and accurate summary. The AST preserves critical structural information: \texttt{MethodDeclaration\_format} identifies the primary action, \texttt{ReferenceType\_String} indicates string creation, \texttt{ReturnStatement} confirms value production, and \texttt{MethodInvocation} suggests processing steps. Despite abstracting away specific identifiers, the AST captures the method's control flow (conditional checks, iteration, and return) and type hierarchy (String return type and LoggingEvent parameter), which in Java's explicit type system carry substantial semantic weight. The model successfully reconstructs accurate functionality from this structural skeleton.


These examples crystallize the quantitative patterns observed across our evaluation: Function Signature's success in Python stems from capturing identifier-level semantics that encode functional intent through descriptive naming conventions, whereas both CrystalBLEU and AST fail by removing complementary but essential information---CrystalBLEU eliminates structural keywords while AST discards semantic identifiers. Full code tokens, paradoxically, perform poorly despite having complete information, suggesting that token density and verbosity can obscure rather than clarify functional intent.
Conversely, in Java, AST excels by preserving structural information that carries substantial semantic weight in Java's explicit type system, while Function Signatures and CrystalBLEU lose critical context by discarding type relationships and control structures. Full code tokens succeed because Java's verbose syntax provides sufficient contextual clarity. This language-specific pattern--where identifiers carry more semantic weight in Python while structural information carries more weight in Java---illustrates why optimal token optimization strategies differ fundamentally between the two languages.

\section{Implications}
\label{sec:implications}
Our findings yield several important implications for code summarization research and practice, spanning representation design, cross-language generalization, and data-centric optimization strategies.
The contrast with our Java results--where AST consistently dominated across all metrics--reveals fundamental differences in how languages encode semantic information. Java's verbose syntax and structured type systems mean that AST representations capture class hierarchies, type declarations, and explicit interfaces that carry substantial semantic weight. Python's concise syntax and dynamic typing shift semantic density toward identifier names and parameter choices, making lexical preservation more valuable than structural abstraction. This language-specificity, observed consistently across our quantitative evaluation, challenges the widespread assumption that syntactic representations like AST provide universal, task-agnostic code understanding.

The broader implication extends to code representation design for neural models. Our results demonstrate that optimal representations must align with both language characteristics and task requirements: structure-preserving representations (AST) suit verbose, explicitly-typed languages and structural tasks (bug detection, type inference), while identifier-preserving representations (Function Signatures) suit concise, descriptively-named languages and semantic tasks (summarization, intent inference). Practitioners deploying code models across languages must validate representation effectiveness per language rather than assuming transferability.

Furthermore, our findings highlight why aggressive token reduction can sometimes outperform full code inputs: by eliminating verbose implementation details and forcing models to rely on concentrated semantic signals (function names, parameters), compression can improve signal-to-noise ratios in training data. This suggests that token optimization serves dual purposes--computational efficiency and semantic focusing--particularly valuable when dealing with noisy or verbose codebases where implementation verbosity obscures functional intent. The observation that semantic filtering enables more aggressive token compression without quality loss further underscores the importance of multi-level data optimization: instance-level filtering (removing low-quality pairs) creates conditions under which token-level compression becomes viable, demonstrating that these optimization dimensions can offer complementary benefits.

Finally, our cross-language analysis signals that data-centric optimization strategies cannot be developed in isolation from language characteristics. The complete reversal of optimal strategies between Java and Python--where the best-performing method in one language becomes the worst in another--demonstrates that ``one-size-fits-all'' optimization is fundamentally misguided. Future work in code summarization and related tasks must adopt language-aware evaluation frameworks that validate techniques across diverse programming paradigms rather than generalizing from single-language studies. This principle extends beyond token optimization to broader questions of model architecture, training objectives, and evaluation metrics, all of which may require language-specific calibration to achieve optimal results.

\section{Threats to Validity}\label{sec:threats}

\textbf{Construct validity.} Our conclusions rely on proxy measures for ``summary quality'' (BLEU, ROUGE-L, METEOR), on \side/\sidep for ``semantic alignment,'' and on token retention and Shannon entropy as proxies for computational efficiency. While these metrics are standard in prior work, they may not capture all facets of summary quality. \side is trained and validated on specific corpora; if its embedding space encodes dataset-specific regularities, \sidep scores may partially reflect distributional artifacts rather than genuine code-summary alignment. Our efficiency measures (token retention, entropy reduction) correlate with computational cost but do not directly measure wall-clock training time or energy consumption, which depend on implementation details (hardware, batch sizes, kernel efficiencies). Finally, contamination between pretraining and evaluation sets can inflate scores, and human assessments (used to validate \sidep and in our qualitative analysis) may carry bias. We applied standard precautions, such as using multiple independent assessors, to mitigate these risks, but we cannot fully exclude them.

\textbf{Internal validity.} Observed differences could be influenced by confounders unrelated to our data-centric interventions. Parser choices for AST/signature extraction (\eg how \texttt{javalang} and \texttt{ast} module handle corner cases) may introduce language-specific biases. The construction of the CrystalBLEU frequent $n$-gram list from the training split introduces potential circularity; any leakage from validation/test into that list would bias results. Thresholds used with \side/\sidep (0.9 in our experiments) can introduce discontinuities around the decision boundary. As with any training pipeline, factors such as random seeds, checkpoint selection, or other hyperparameter choices may contribute residual variance. We controlled for these factors by using identical training configurations across all conditions. However, some confounding effects may still remain.

\textbf{Conclusion validity.} We address multiple comparisons by applying Holm-Bonferroni correction to p-values and complement statistical significance tests with effect size measures (Cliff's Delta). We compare many conditions (raw vs.\ SIDE-filtered; three token-level strategies; two languages; two evaluation benchmarks), increasing the risk of false positives. Several metrics are correlated but not interchangeable, and some effect sizes are modest even when $p$-values are small. We therefore emphasize effect magnitudes and cross-metric patterns rather than single-metric significance alone. All statistical results and raw data are available in our replication package, enabling independent verification.

\textbf{External validity.} Generalization may be limited along several dimensions: (i) \textit{Programming languages}---we evaluated only Java and Python. Other languages (\eg C/C++, JavaScript, Rust, Go) with different syntax characteristics, naming conventions, and documentation practices may exhibit different trade-offs. For instance, languages with more verbose syntax or less descriptive naming conventions may favor different optimization strategies. (ii) \textit{Model architectures and sizes}---our experiments used CodeT5+ 220M. Larger models (\eg 770M, 2B, 16B parameters) or different architectures (decoder-only models like GPT, encoder-only models like CodeBERT) may show different sensitivities to token reduction. Larger models might be more robust to aggressive compression, while smaller models might suffer more from information loss. (iii) \textit{Dataset characteristics}---both FuncomJava and FuncomPython were mined from GitHub repositories with specific quality controls. Codebases with different documentation practices, code complexity distributions, or domain characteristics (\eg embedded systems, scientific computing, web applications) may respond differently to token optimization strategies. (iv) \textit{Task variations}---we focused exclusively on method-level code summarization. Other code intelligence tasks (\eg commit message generation, API documentation generation, code search) may benefit from different token representations and optimization strategies. (v) \textit{Evaluation contexts}---our evaluation used specific benchmarks (CoderEval, Mastropaolo \etal, New-Benchmark) with particular characteristics. Performance on industrial codebases with proprietary documentation standards, or on real-time code assistance scenarios with strict latency requirements, may differ from our findings. Despite these limitations, our multi-language, multi-dataset, multi-metric evaluation provides evidence that token optimization strategies exhibit language-dependent effectiveness patterns, which practitioners should validate before deploying them in production systems.

\section{Conclusions and Future Works}
\label{sec:conclusion}

We investigated how token-based optimization strategies affect the quality and efficiency of LLM-generated code summaries. Our results show that \emph{which} tokens we retain matters, and \emph{how many}. Structure- and frequency-aware reductions can substantially lower training compute while preserving---and sometimes improving---summary quality. CrystalBLEU-guided pruning and AST-based representations provide dependable compression with minimal degradation, whereas Function Signatures deliver the most aggressive savings when applied after semantic filtering, making them especially suitable for resource-constrained settings.

\noindent\textbf{Practical recipe.} First, semantically filter misaligned code---comment pairs. Then, pick the representation that better suits the data and the intended optimization goal: CrystalBLEU for robustness on noisier corpora, Function Signatures for maximal savings on clean data, and AST when lexical fidelity to surface tokens is required.

\noindent\textbf{Looking ahead.} Future work should explore instance-adaptive reduction policies, dynamic selection of reductions during training, and evaluation protocols that better capture semantic adequacy beyond surface form.

\section{Data Availability}
\label{sec:da}
Datasets, scripts, and raw results used in this study are publicly available~\cite{replication}.

\section{Acknowledgments}
This research was funded by NSF CCF-2451058. A complete list of image attributions can be found at \cite{replication}.

\bibliographystyle{ACM-Reference-Format}
\bibliography{main}
\end{document}